\documentclass[apj]{emulateapj}
\usepackage{mathptmx}

\newcommand  \kms      {\ifmmode {\rm km\,s}^{-1} \else km\,s$^{-1}$\fi}

\newcommand  \ergs     {\ifmmode {\rm ergs\,s}^{-1} \else ergs s$^{-1}$\fi}
\newcommand  \ergcms   {\ifmmode {\rm ergs\,cm}^{-2}\,{\rm s}^{-1}
                        \else ergs\,cm$^{-2}$\,s$^{-1}$\fi}
\newcommand  \ergcmsA  {\ifmmode{\rm ergs\,cm}^{-2}\,{\rm s}^{-1}\,{\rm\AA}^{-1}
                        \else ergs\,cm$^{-2}$\,s$^{-1}$\,\AA$^{-1}$\fi}
\newcommand  \ergcmsHz {\ifmmode{\rm ergs\,cm}^{-2}\,{\rm s}^{-1}\,{\rm Hz}^{-1}
                        \else ergs\,cm$^{-2}$\,s$^{-1}$\,Hz$^{-1}$\fi}
\newcommand  \phcms    {\ifmmode {\rm ph\,cm}^{-2}\,{\rm s}^{-1}
                        \else ,ph\,cm$^{-2}$\,s$^{-1}$\fi}
\newcommand  \phcmsA   {\ifmmode {\rm ph\,cm}^{-2}\,{\rm s}^{-1}\,{\rm\AA}^{-1}
                        \else ph\,cm$^{-2}$\,s$^{-1}$\,\AA$^{-1}$\fi}      
\def\micron{\ifmmode \mu{\rm m} \else $\mu$m\fi}
\def\kms{\ifmmode {\rm km\,s}^{-1} \else km\,s$^{-1}$\fi}
\def\Hubble{\ifmmode {\rm km\,s}^{-1}\,{\rm Mpc}^{-1}
        \else km\,s$^{-1}$\,Mpc$^{-1}$\fi}
\def\ergsec{\ifmmode {\rm ergs\;s}^{-1} \else ergs s$^{-1}$\fi}
\def\ergscm{\ifmmode {\rm ergs\,s}^{-1}\,{\rm cm}^{-2}
          \else ergs\,s$^{-1}$\,cm$^{-2}$\fi}
\def\ergscmA{\ifmmode {\rm ergs\,s}^{-1}\,{\rm cm}^{-2}\,{\rm \AA}^{-1}
          \else ergs\,s$^{-1}$\,cm$^{-2}$\,\AA$^{-1}$\fi}
\def\ergscmHz{\ifmmode {\rm ergs\,s}^{-1}\,{\rm cm}^{-2}\,{\rm Hz}^{-1}
          \else ergs\,s$^{-1}$\,cm$^{-2}$\,Hz$^{-1}$\fi}
\def\Msun{\ifmmode M_{\odot} \else $M_{\odot}$\fi}
\def\Lsun{\ifmmode L_{\odot} \else $L_{\odot}$\fi}
\def\qo{\ifmmode q_{0} \else $q_{0}$\fi}
\def\Ho{\ifmmode H_{0} \else $H_{0}$\fi}
\def\ho{\ifmmode h_{0} \else $h_{0}$\fi}
\def\qo{\ifmmode q_{0} \else $q_{0}$\fi}
\def\ao{\ifmmode a_{0} \else $a_{0}$\fi}
\def\to{\ifmmode t_{0} \else $t_{0}$\fi}
\def\Halpha{\ifmmode {\rm H}\alpha \else H$\alpha$\fi}
\def\Hbeta{\ifmmode {\rm H}\beta \else H$\beta$\fi}
\def\hb{\ifmmode {\rm H}\beta \else H$\beta$\fi}
\def\Hgamma{\ifmmode {\rm H}\gamma \else H$\gamma$\fi}
\def\Hdelta{\ifmmode {\rm H}\delta \else H$\delta$\fi}
\def\Lya{\ifmmode {\rm Ly}\alpha \else Ly$\alpha$\fi}
\def\Lyb{\ifmmode {\rm Ly}\beta \else Ly$\beta$\fi}
\def\hi{\ifmmode \mbox{{\rm H}\,{\sc i}} \else H\,{\sc i}\fi}

\def\ciii{\ifmmode {\rm C}\,{\sc iii} \else C\,{\sc iii}\fi}

\def\o5007{[O\,{\sc iii}]\,$\lambda5007$}
\def  \kms         {\hbox{km s$^{-1}$}}          
\def  \ergs        {\hbox{erg s$^{-1}$}}              

\def  \etal        {{\rm et al.}}
\def  \La          {\ifmmode {\rm Ly}\alpha \else Ly$\alpha$\fi}
\def  \Ka          {\ifmmode {\rm K}\alpha \else K$\alpha$\fi}
\def  \Lb          {\ifmmode {\rm L}\beta \else L$\beta$\fi}
\def  \Ha          {\ifmmode {\rm H}\alpha \else H$\alpha$\fi}
\def  \Hb          {\ifmmode {\rm H}\beta \else H$\beta$\fi}
\def  \Pa          {\ifmmode {\rm P}\alpha \else P$\alpha$\fi}
\def  \CIIIb       {\ifmmode {\rm C}\,{\sc iii]}\,\lambda1909
                     \else C\,{\sc iii]}\,$\lambda1909$\fi}
\def  \CIV         {\ifmmode {\rm C}\,{\sc iv}\,\lambda1549
                     \else C\,{\sc iv}\,$\lambda1549$\fi}
\def  \MgII         {\ifmmode {\rm Mg}\,{\sc ii}\,\lambda2798
                     \else Mg\,{\sc ii}\,$\lambda2798$\fi}
\def  \OVI         {\ifmmode {\rm O}\,{\sc vi}\,\lambda1035
x
                     \else O\,{\sc vi}\,$\lambda1035$\fi}
\def \chandra  {{\it Chandra}}
\def \xmm      {{\it XMM-Newton}}

\journalinfo{The Astrophysical Journal, astro-ph/0309096}
\slugcomment{Received 2003 July 7; accepted 2003 September 1}

\shorttitle{VARIABILITY AND MODELING OF NGC\,3783}
\shortauthors{NETZER ET AL.}

\begin{document}
 
\title{The Ionized Gas and Nuclear Environment in NGC\,3783. \\
  IV. Variability and Modeling of the 900 ks {\it CHANDRA} Spectrum}

\author{
Hagai Netzer,\altaffilmark{1}
Shai Kaspi,\altaffilmark{1} 
Ehud Behar,\altaffilmark{2}
W. N. Brandt,\altaffilmark{3} 
Doron Chelouche,\altaffilmark{1} 
Ian M. George,\altaffilmark{4,5}
D. Michael Crenshaw,\altaffilmark{6}
Jack R. Gabel,\altaffilmark{7}
Frederick W. Hamann,\altaffilmark{8}
Steven B. Kraemer,\altaffilmark{7}
Gerard A. Kriss,\altaffilmark{9,10}
Kirpal Nandra,\altaffilmark{11}
Bradley M. Peterson,\altaffilmark{12}
Joseph C. Shields,\altaffilmark{13}
and 
T. J. Turner,\altaffilmark{4,5}
}

\altaffiltext{1}{School of Physics and Astronomy, Raymond and Beverly Sackler
Faculty of Exact Sciences, Tel-Aviv University, Tel-Aviv 69978, Israel.}
\altaffiltext{2}{Physics Department, Technion, Haifa 32000, Israel}
\altaffiltext{3}{Department of Astronomy and Astrophysics, 525 Davey
Laboratory, The Pennsylvania State University, University Park, PA 16802.}
\altaffiltext{4}{Laboratory for High Energy Astrophysics, NASA/Goddard Space
Flight Center, Code 662, Greenbelt, MD 20771.}
\altaffiltext{5}{Joint Center for Astrophysics, Physics Department, University
of Maryland, Baltimore County, 1000 Hilltop Circle, Baltimore, MD 21250.}
\altaffiltext{6}{Department of Physics and Astronomy, Georgia State
University, Atlanta, GA 30303.}
\altaffiltext{7}{Catholic University of America, NASA/GSFC, Code 681,
Greenbelt, MD 20771.} 
\altaffiltext{8}{Department of Astronomy, University of Florida, 211 Bryant 
Space Science Center, Gainesville, FL 32611-2055.}
\altaffiltext{9}{Center for Astrophysical Sciences, Department of Physics and
Astronomy, The Johns Hopkins University, Baltimore, MD 21218-2686.}
\altaffiltext{10}{Space Telescope Science Institute, 3700 San Martin Drive, Baltimore,
  MD 21218}
\altaffiltext{11}{Astrophysics Group, Imperial College London, iBlackett Laboratory,
Prince Consort Rd., London SW7 2AZ, UK }
\altaffiltext{12}{Department of Astronomy, Ohio State University, 140 West 18th
Avenue, Columbus, OH 43210-1106.}
\altaffiltext{13}{Department of Physics and Astronomy, Clippinger
Research Labs 251B, Ohio University, Athens, OH 45701-2979.}

\begin{abstract}                    

We present a detailed spectral analysis of the data obtained from NGC\,3783 
during the period 2000--2001 using \chandra.
The data were split in various ways to look for time- and luminosity-dependent  
spectral variations. This analysis, along with the measured equivalent widths of a large 
number of  X-ray lines and photoionization calculations, lead us to
the following results and conclusions.
1) NGC\,3783 fluctuated in luminosity by a factor $\sim 1.5$ during individual
observations (most of which were of 170 ks duration). 
These fluctuations were not associated with significant spectral variations.
2) On a longer time scale (20--120 days), we found the source 
to exhibit two very different spectral shapes. 
The main difference between these  can be well-described by 
the appearance (in the  ``high state'') and disappearance (in the
``low state'') of a spectral component that dominates the 
underlying continuum at the longest wavelengths.
Contrary to the case in other objects, the spectral variations are not related 
to the brightening or the fading of the continuum at short wavelengths
in any simple way.
NGC\,3783 seems to be the first AGN to show this unusual behavior.
3) The appearance of the soft continuum component is consistent with being
{\it the only} spectral variation, and there is no
need to invoke changes in the opacity of the absorbers lying along the line of sight.
Indeed, we find all the absorption lines which can be reliably measured
have the same equivalent widths (within the observational uncertainties)
 during high- and low-states.
4) Photoionization modeling indicates that a combination of three 
ionized absorbers, each split into two kinematic components, can explain the
strengths of almost all the absorption lines and bound-free edges.
These three components span a large range of ionization, and have
total column of about 4$\times 10^{22}$ cm$^{-2}$.
Moreover, all three components are thermally stable and seem to have the same gas
pressure. Thus all three
may co-exist in the same volume of space.
This is the first detection of such a multi-component, equilibrium gas in an AGN. 
5) The only real discrepancy between our model and the observations 
concerns the range of wavelengths absorbed by the iron M-shell UTA feature. 
This most likely arises as the result of our underestimation of the 
poorly-known dielectronic recombination rates appropriate for these ions.
We also note a small discrepancy in the calculated column density of \ion{O}{6} and
discuss its possible origin.
6) The lower limit on the distance of the absorbing gas in NGC\,3783 is between
0.2 and 3.2 pc, depending on the component of ionized gas considered.
The assumption of pressure equilibrium imposes an upper limit of about 25 pc on the 
distance  of the least-ionized component from the central source.

\end{abstract}


\section{Introduction}

The barred-spiral galaxy 
NGC\,3783 ($V \simeq 13.5$ mag., $z=0.0097$) hosts a 
well-studied, type-I  active galactic nucleus (AGN)
with prominent broad emission lines and strong X-ray absorption features.
The object has been observed extensively with almost
all X-ray instruments, most
recently by \chandra\ (Kaspi et al. 2002, hereafter Paper I) and \xmm\ (Blustin et al 2002).
The 2--10 keV flux of NGC\,3783 varies in the range 
$\sim(4$--$9)\times 10^{-11}$~\ergcms , and its mean 2--10 keV luminosity is
$\sim 1.5\times 10^{43}$~\ergs\ (for $H_0=70$~km\,s$^{-1}$\,Mpc$^{-1}$ and
$q_0=0.5$). 
Paper I gives an extensive list of references and a comprehensive summary 
of historical observations, including ground-based and UV
(HST) observations.  It also discusses the unique 
\chandra\ data set obtained in 2000--2001. 
These observations consist of a relatively short
        observation performed in 2000 January, and five longer observations
        performed in 2001 February -- June, separated by various intervals from
        2 to 120 days (see \S2).
Paper~I contains numerous
illustrations of the mean spectrum, absorption line profiles  and 
detailed measurements of many absorption and emission lines.
Two other papers (Gabel et al. 2003a,b) discuss the 
{\it HST} and {\it FUSE} data obtained as part of this
multi-waveband campaign.

There have been several previous attempts to model the 
characteristics of the X-ray absorbing gas along the line of sight to  NGC\,3738.
Here we comment only on the more detailed works.
A very detailed work Kaspi et al
(2001) attempted to fit the spectrum obtained with \chandra\ in 2000.
The Kaspi et al.  most successful model consists of two
absorbing  shells, both outflowing from the central source
with a velocity of $\sim 600$ \kms. The gas in both shells 
was assumed to be turbulent with internal turbulent motion of $\sim 250$ \kms.
The two shells had similar column densities, but different ionization parameters. 
The model is consistent with the intensity 
and equivalent width (EW) of many (but not all) emission and 
absorption lines observed in the spectrum.
A major limitation of the Kaspi et al. (2001) work was 
the limited signal-to-noise (S/N) of the spectrum, which  
results in large uncertainties on the model parameters. 
In addition, the limited duration of the only observation then available
did not allow a meaningful analysis of any time-dependence.
 A new paper, 
by Krongold et al. (2003, hereafter K03), discussing  
 the full \chandra\ data set, was accepted for publication
after the submission of our paper. Some results of that work are relevant
to our study and are discussed in the various sections below. 

This paper discusses the complete spectroscopic \chandra\ data of NGC\,3783.
We present 
our modeling of this material, with emphasis on the properties of the 
absorber(s) along the line of sight.
In \S2 we describe the two X-ray spectral states discovered using the new 
observations.
In \S3 we explain the various ways we measured and modeled the spectrum.
\S4 contains a discussion of the new findings, and  \S5 summarizes our
new results.

\section{The two-state X-ray spectrum of NGC\,3783}

A full description of the \chandra\ data considered here is given in Paper I.
In brief, there were 5 observations (each of $\sim$170~ks duration) 
obtained over the period 2001 February -- June, and a 
shorter observation (of $\sim$56~ks duration) obtained in 2000 January
(and also discussed in Kaspi et al. 2001).
All six observations were performed with
the High-Energy Transmission Grating Spectrometer (HETGS) 
in place, and consist of a total exposure of 888.7 ks. 
All measurements in Paper I refer to a time-averaged spectrum,
produced by combining the first-order spectra 
from both grating arms,  using 0.01\AA\ wide bins. 
Third-order data from the medium-energy grating (MEG)  were also used to compare
the profiles of several short-wavelength lines with the first-order profiles obtained for
several lines at longer wavelength. 
It was found that most of the resolved absorption
lines (e.g. \ion{O}{7}, \ion{Ne}{9} and \ion{Si}{14} lines) 
consist of at least two kinematic
components, outflowing at $-500 \pm 100$ \kms\ and 
$-1000 \pm 200$ \kms\ (respectively).
The overall absorption profiles in NGC\,3783 
covers the velocity range of 0 to -1600 \kms\ (e.g. see Figs. 5 \& 10 in Paper I).

A major goal of the present work is to investigate subsets of the \chandra\ data.
We subdivided the data in various ways 
to search for spectral variations as a function of source luminosity and/or time.
This was done for both short  (i.e. within the individual 170 ks observations)
as well as for long (between observations i.e. 20--90 days) time
scales.
The short time-scale behavior of NGC\,3783 is illustrated in Fig.1 
which shows the short wavelength (2--10~\AA) flux as a function of time
for all observations binned in intervals of 3170 sec in the left panels. The right panels 
shows the "softness" ratio, defined as the flux in the 15--25\AA\ band divided
by that in the 2--10~\AA\ band, to illustrate
now the different variability characteristics exhibited by the source 
above and below $\sim$1~keV.
\begin{figure*}[t]
\centerline{\includegraphics[width=17.2cm]{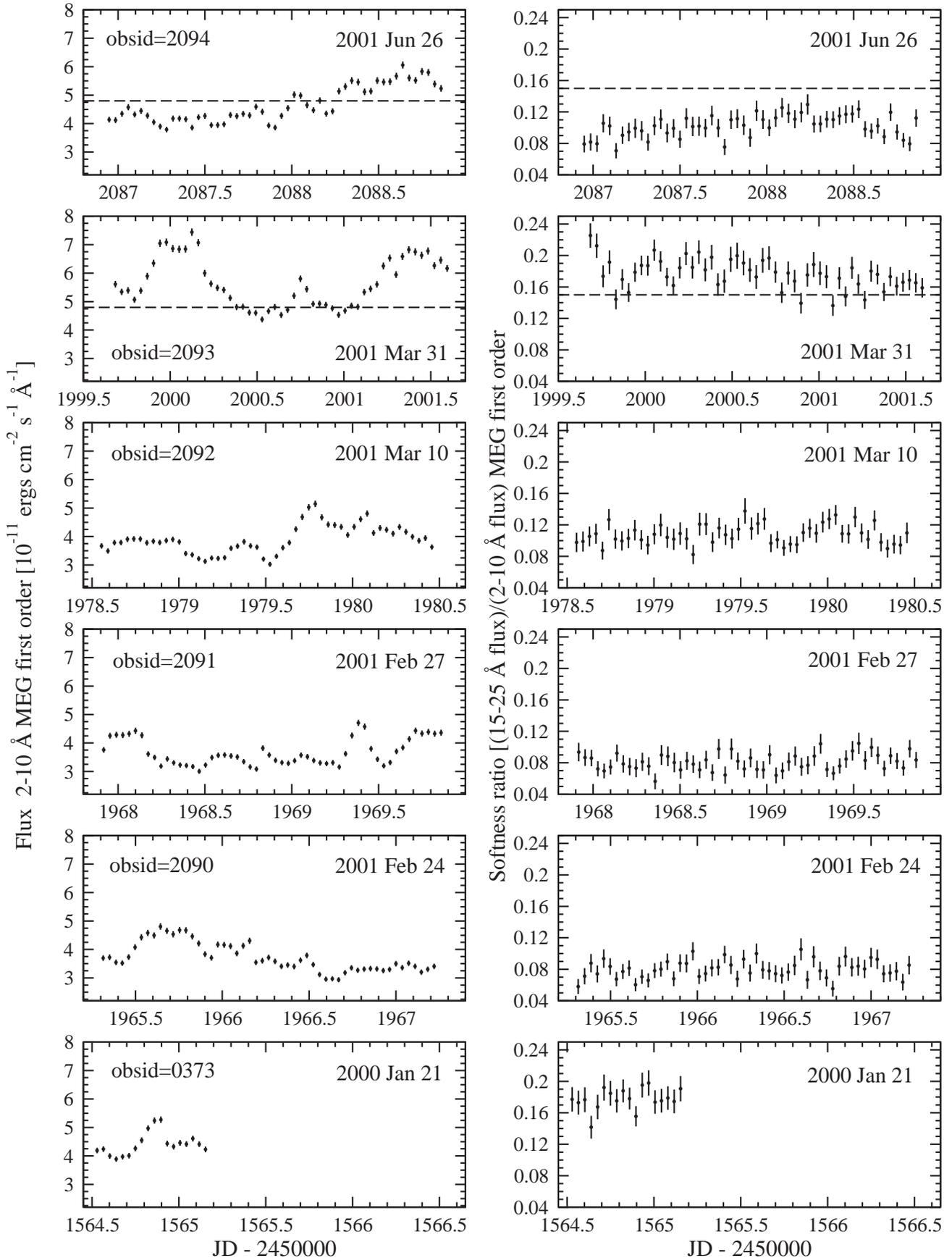}}
\caption{MEG 2--10~\AA\ flux light curves (left)
and  softness ratios (ratio of the 2--10 and the 15--25~\AA\ count
rates,  right) for all \chandra\ observations of NGC\,3783.
The dashed lines across observations no. 2093 and 2094 represent constant
flux (4.8) and softness ratio (0.15). They
illustrate the fact that similar 2--10~\AA\ fluxes (middle part of obs.
2093 and last part of obs. 2094) can be associated with very different
softness ratios.
\label{fig1_label} }
\end{figure*}

From Fig. 1 it can be seen that the  2--10~\AA\ flux varied by about 50\% over the 170 ks 
duration of each of the observations. However,  
inspection of the corresponding softness light curves shows that the 
spectral variations are much smaller and almost insignificant. 
 Thus, the data show very little cahnges of the spectral energy distribution (SED)  over 
 time scales $\lesssim$4~days ("obsids" 2090 and 2091 are consecutive with total
duration of about 4 days). To further test this finding, we have also
extracted various spectra when the source was in different intensity states, 
but did not find any significant spectral variations during this period.
We also note that Behar et al. (2003) did not find any 
spectral variations during a three-day observation of 
NGC 3783 in 2001 December using \xmm.

Contrary to the above, there is a {\it very significant} change of SED on 
longer time scales. In particular, there is a major change 
in the softness ratio between obsids 2092 and  2093
(separated by 20 days) and obsids 2093 and 2094 
(separated by 90 days). As shown in Fig. 1, obsid 2093
shows a higher mean count rate but exhibits a much softer spectrum. Direct
comparison of  the time-averaged spectra obtained during 
obsid  2093 with that obtained  during obsid 2094 
reveals the low energy part ($\lambda > 15$\AA) decreased  in flux by
a  factor of $\sim 4$, but that the shorter wavelength continuum exhibited 
a much smaller decrease (factor $\sim 1.5$)
The opposite change occured between obsids 
2092  and  2093.  Most importantly, the softness ratio
variations {\it are not} simple luminosity-related effects. 
For example, the
2--10~\AA\ count rate in the middle of obsid 2093 (Fig. 1) is a
little lower than the 2--10~\AA\ flux toward the end of obsid
2094, yet the softness ratios are significantly different.
A similar effect is seen in obsid 2093, during which the
2--10~\AA\ count rate varied by about 50\% yet the
softness ratio remaining approximately constant.
 
In light of these results, we have divided the entire data set into 
groups with high and low softness ratios. We find 
four observations (2090, 2091, 2092 and 2094) with low softness ratio
and two (0373 and 2093) with high softness ratio.
Hereafter we refer to these as the "low-" and "high-state"
          observations, respectively. 
We find no significant differences between the mean spectra
       of the four individual low-state observations (except for a small
       intensity variations). Similarly we find no significant differences
       between the mean spectral shapes of the two high-state observations.

Fig. 2 is a more detailed example of this phenomenon. It shows the softness ratio as
a function of the 2--10~\AA\ flux  for all the data using 3170s bins. A separation into two
groups is apparent,  and a standard KS statistical test confirms its significance. We have also
investigated possible linear correlations of the softness ratio with
the  2--10~\AA\ flux within each of the states. 
For the low-state observations, we find a significant ($>99$\%) linear correlation between softness ratio and flux.
However, as can be seen from Fig.2, clearly this line does not connect the low- and 
high-state data. 
No significant correlation of softness ratio with flux was found for the high-state data.
We conclude that the X-ray spectrum of NGC\,3783 fluctuates between two states of
different softness ratio.  The combined spectra of the two are shown in Fig. 3 and
much of the rest of this paper is devoted to the analysis of this unusual behavior of NGC\,3783.  
\begin{figure}
\centerline{\includegraphics[width=8.5cm]{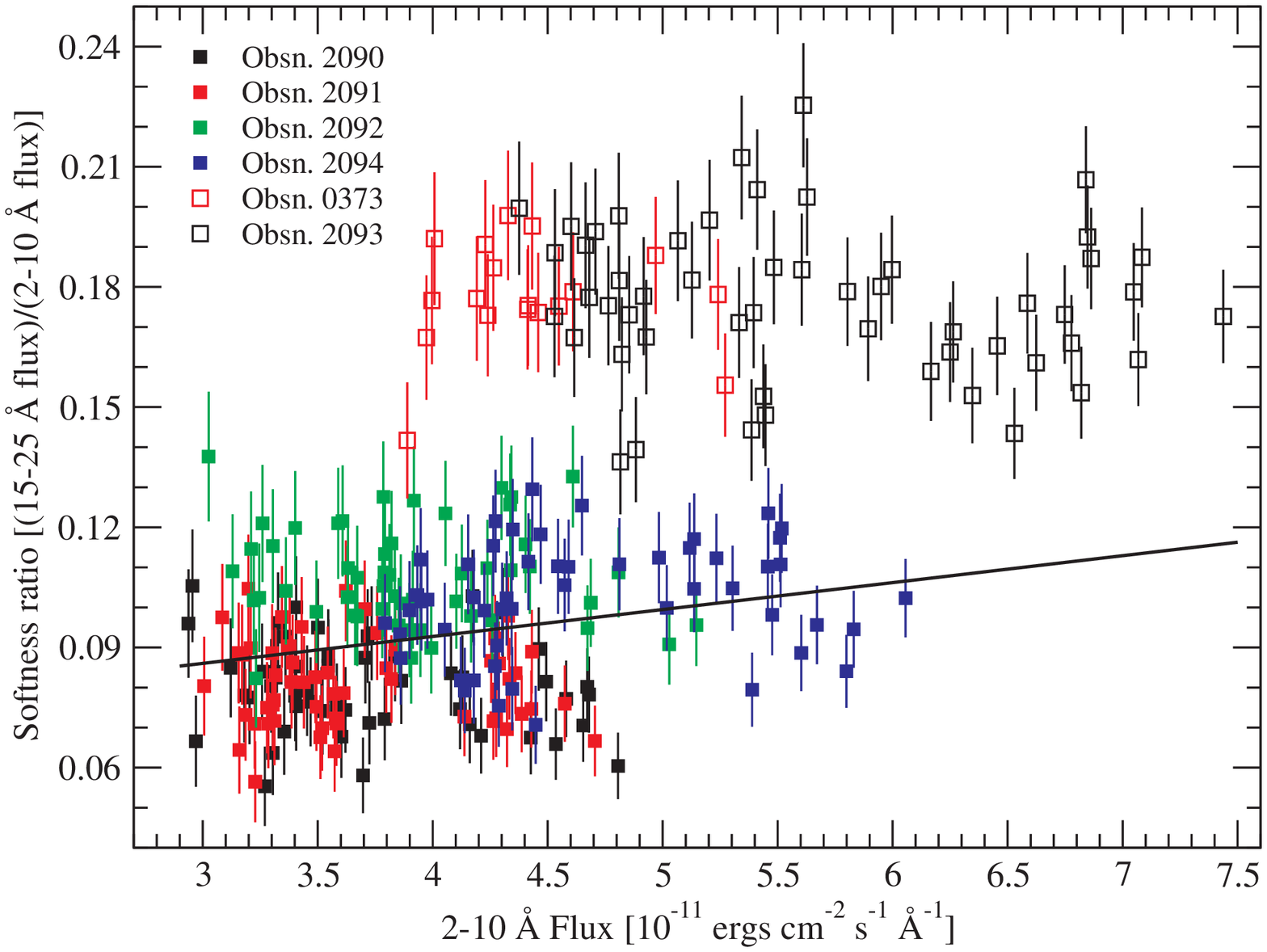}}
\figcaption {Softness ratio vs. 2--10~\AA\ flux for the data in Fig. 1.
The low and high-state observations are shown
        as filled and open symbols, respectively, and the various observations are
denoted with different colors.  The solid straight line is the linear regression fit to the
        low-state observations. No significant
correlation of softness ratio vs. flux was found for the high state observations.
\label{fig2_label} }
\end{figure}

\begin{figure}
\centerline{\includegraphics[width=10cm]{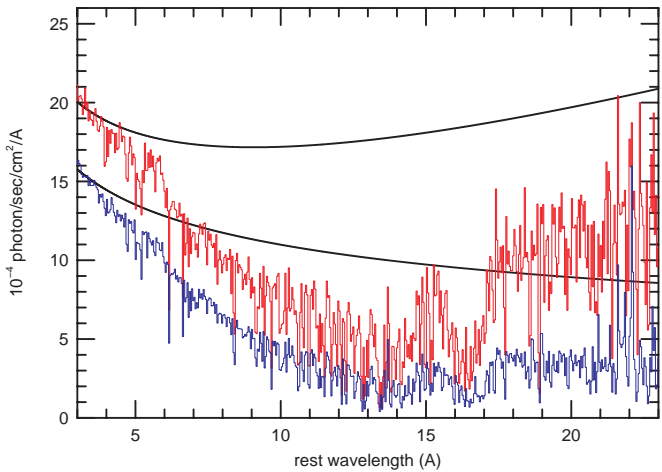}}
\figcaption {High-state (red, observations 0373 and 2093 combined) and low-state (blue, all other
observations combined) spectra of NGC\,3783 binned to 0.04~\AA.
The solid lines are the two chosen intrinsic continua (see \S3.2.5). The lower continuum is a single
power-law with $\Gamma=1.6$. The high continuum is made out of two
power-laws with slopes $\Gamma=1.6$ and $\Gamma=3.2$ and relative normalization
of 1:0.4 at 1 keV. These power-laws are discussed in \S3.2.5.
\label{fig3_label} }
\end{figure}

\section{Spectral analysis and modeling}

\subsection{Spectral differences between the high and the low states}

We have investigated the unusual variations observed in the spectrum of NGC\,3783
in an  attempt to understand their nature and their origin. In particular,
we have attempted to answer the question of whether they are due to the response of the 
absorbing gas  to the variations observed in the continuum. 

There are two ways to answer the above question -- by a direct and detailed spectral
comparison, and by modeling the two spectra trying to establish the origin of the differences.
Fig. 4 is an example of the first approach.  The diagrams shows a
comparison of the low- and high-state spectra after applying a simple 
scaling factor to the former such that the (local) continua have the same intensity.
As the scaling factor is wavelength dependent, we divided
the two spectra into several  bands and applied different factors in each case.
The two examples shown in Fig. 4 (and all others we have examined) illustrate that,
except for a luminosity scaling, the two absorption line spectra are
indistinguishable within the uncertainties ($\sim$10\% for the low-state spectrum and
about twice that for the high-state spectrum).
The somewhat weaker looking lines of low ionization
species in the high-state spectrum (e.g. around 7\AA) are well within the noise.
Table 1 lists the EWs measured for several key absorption lines (see \S3.2.3) in 
both the high- and low-state spectra. 
Again, the small differences between the two spectra are
well within the measurement errors. 
We proceed under the assumption that there are no variations in the line EWs between the two states. We comment on the implications of this in \S4.1.

\begin{figure*}
\hglue-0.4cm{\includegraphics[angle=0,width=10cm]{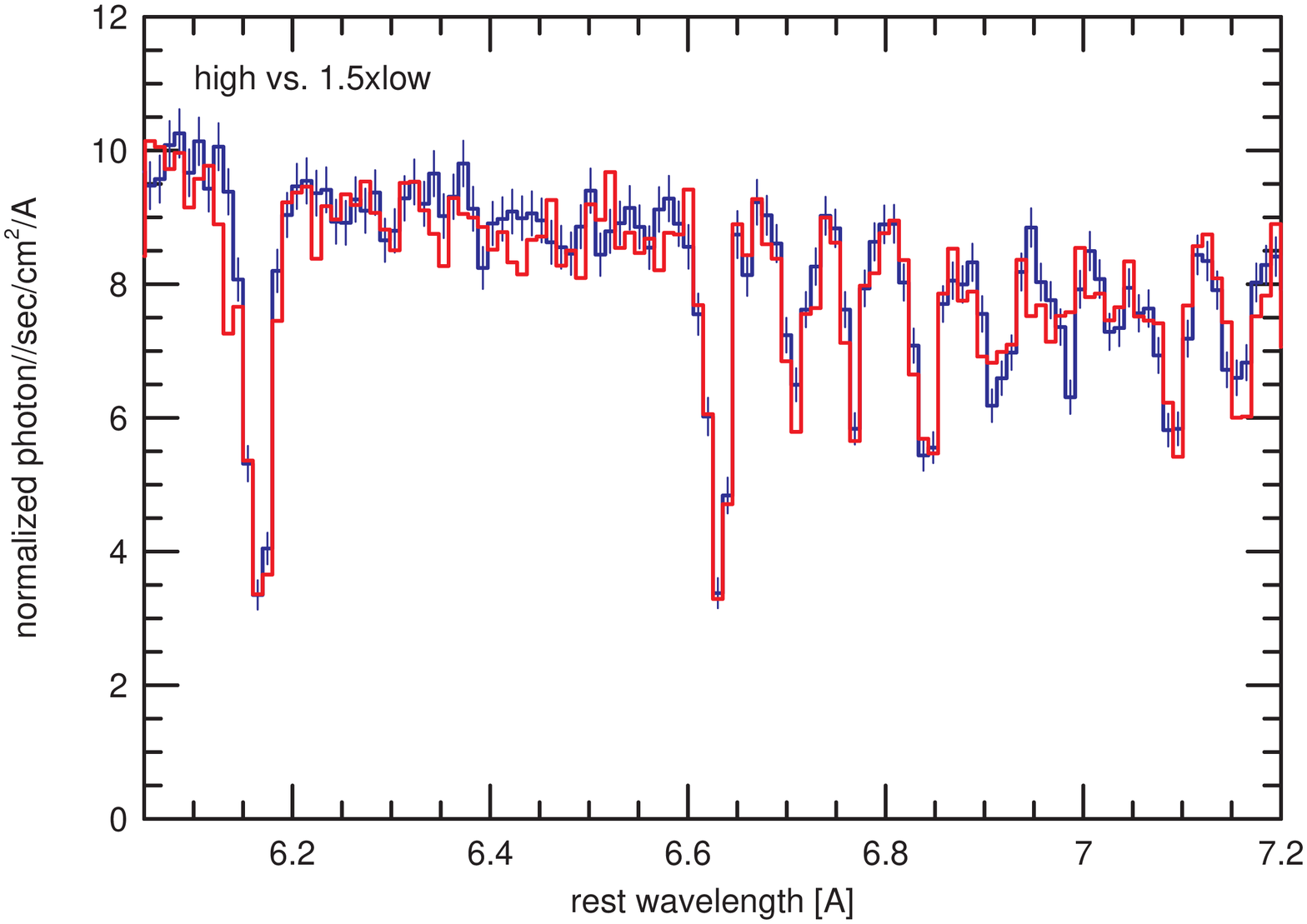}}
\hglue-0.8cm{\includegraphics[angle=0,width=10cm]{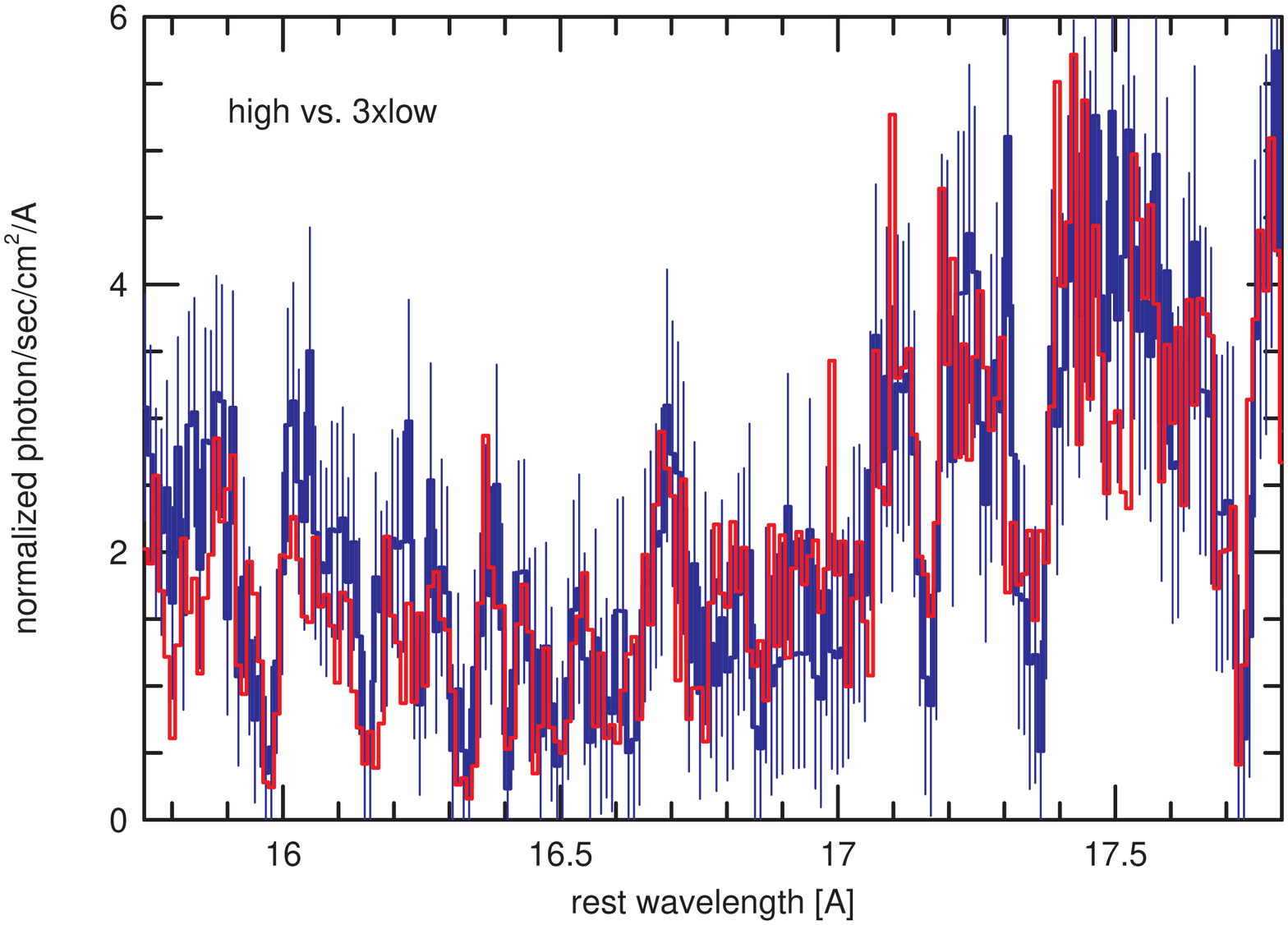}}
\caption{Left: High-state (red) vs. low-state (blue) 6--7~\AA\ spectra of NGC\,3783 showing
the great similarity in absorption line EWs (cf. table 1). The low-state spectrum was
multiplied by 1.5 to match the flux of the  high-state continuum.
For clarity, error bars are only plotted for the low state spectrum.
Right: Same for the 15.6--18~\AA\ range except for the
different scaling of the low-state continuum. }
\label{fig4_label}
\end{figure*}

\begin{deluxetable}{lccc}
\tablecolumns{4}
\tablewidth{0pt}
\tablecaption{Equivalent Widths\tablenotemark{a}
\label{ews}}
\tablehead{
\colhead{Ion \& Line} &
\colhead{low state} &
\colhead{high state} &
\colhead{Kaspi et al. (2002)} }
\startdata
\ion{Si}{14} $\lambda 6.182$ & $20.4\pm1.2$ & $22.6\pm1.7$ & $20.5\pm0.8$ \\
\ion{Si}{13} $\lambda 6.648$ & $16.0\pm1.3$ & $14.6\pm1.3$ & $14.9\pm0.7$ \\
\ion{Si}{12} $\lambda 6.718$\,\tablenotemark{b} & \phn $2.9\pm0.5$ & \phn $3.8\pm0.7$ & \phn $3.0\pm0.4$ \\
\ion{Si}{11} $\lambda 6.778$ & \phn $4.8\pm0.8$ & \phn $5.9\pm1.2$ & \phn $4.8\pm0.6$ \\
\ion{Si}{10} $\lambda 6.859$ & $10.6\pm1.1$ & \phn $9.4\pm1.4$ & $10.9\pm0.8$ \\
\ion{Si}{9} $\lambda  6.931$ & \phn $7.2\pm1.0$ & \phn $7.7\pm1.6$ & \phn $7.1\pm0.7$ \\
\ion{Si}{8} $\lambda  6.999$ & \phn $4.3\pm1.0$ & \phn $4.4\pm1.4$ & \phn $4.3\pm0.8$ \\
\ion{S}{16} $\lambda  4.729$ & $11.5\pm1.6$ & $11.6\pm2.5$ & $10.7\pm1.2$ \\
\ion{S}{15} $\lambda  5.039$ & \phn $8.7\pm1.7$ & \phn $9.1\pm2.5$ & \phn $9.2\pm1.2$ \\
\enddata
\tablenotetext{a}{Equivalent widths were measured as described in
Paper I and are given in m\AA .}
\tablenotetext{b}{This line is blended with \ion{Mg}{12}\,$\lambda
6.738$ with 1:1 ratio. Tabulated values are only for the
\ion{Si}{12}\,$\lambda 6.718$ line.}
\end{deluxetable}

\subsection{Spectral modeling}
The second and complementary approach for investigating the spectral changes is by
detailed modeling of the absorbing and emitting gas.
 The underlying idea is that the gas is photoionized by the
central X-ray source, and that the observed spectra represent its physical state
during the two states. The principles and the ingredients of such modeling were outlined in
Netzer (1996) and previous applications to the case of NGC\,3783 were discussed
by  Kaspi et al. (2001). 
 A very recent analysis is provided by K03 who 
describe a detailed model composed of two absorbers that are different
from those suggested by Kaspi et al. (2001).
Below we comment on the similarity and differences
 between our work and the new  K03 paper.
Here we summarize our basic method and explain its
application to the high and the low-state spectra of NGC\,3783.
 
\subsubsection{General method and model ingredients}
The X-ray gas is assumed to be photoionized by a central source, and 
in  photoionization and thermal equilibrium (see \S4 for discussion of the last
point). Modeling was performed using ION2003, the 2003 version of the code ION
(Netzer 1996). The code includes all relevant atomic 
processes, and computes the ionization and thermal structure of the gas
along with the intensities and EWs for more than 2000 X-ray lines. 
The code is able to consider various geometries, from a single cloud to a
multi-component, expanding atmosphere. 
The basic parameters of the model are
the gas density (assumed to be in the range 10$^{2-6}$ cm$^{-3}$ - see
justification in \S4), the hydrogen column density ($N$, in units of cm$^{-2}$), 
the covering factor,
 the gas composition, turbulent motion (\S3.2.2) and
the oxygen ionization parameter ($U_{OX}$) defined over the energy range
0.54--10 keV. As explained by Netzer (1996), and discussed further in George
et al. (1998), this choice of ionization parameter (compared with, for example, 
$U_X$ defined over the 0.1--10 keV range) gives the most meaningful definition of the 
ionization field of X-ray photons for AGN. Regarding the covering factor, 
we distinguish between the emission covering factor ($\Omega /4 \pi$) applicable to
the emission line gas, and the absorption (line of sight) covering factor which can be 
different.  The ``solar composition'' used throughout this work is given in Table 2
(note the reduced oxygen abundance compared with older estimates).

\begin{deluxetable}{cc}
\tablecolumns{2}
\tablewidth{0pt}
\tablecaption{Assumed composition
\label{Element}}
\tablehead{
\colhead{Element  } &
\colhead{Relative abundance }
}
\startdata
H & 1.0 \\
He  & 0.1 \\
C  & $ 3.7 \times 10^{-4}$ \\
N  & $ 1.1 \times 10^{-4}$ \\
O  & $ 5.0 \times 10^{-4}$ \\
Ne  & $ 1.0 \times 10^{-4}$ \\
Mg  & $ 3.7 \times 10^{-5}$ \\
Al  & $ 3.0 \times 10^{-6}$ \\
Si  & $ 3.5 \times 10^{-5}$ \\
S  & $ 1.6 \times 10^{-5}$ \\
Ar  & $ 3.3 \times 10^{-6}$ \\
Ca  & $ 2.3 \times 10^{-6}$ \\
Fe  & $ 4.0 \times 10^{-5}$ \\

\enddata
\end{deluxetable}

The incident continuum is taken to be the broken power-law
defined in Kaspi et al. (2001, Table 4). The only changes we have experimented
with are related to the slope of the 0.1--50 keV continuum. As discussed below, the UV part 
of the spectrum can be different from the one assumed here with important 
implications for the UV absorption lines. This will be investigated in a forthcoming paper.
The effects of the UV continuum on the strongest features seen in the X-ray band are of
far less importance. Thus we consider this SED to be adequate for the present
analysis.

The models calculated here are entirely self-consistent and are not simple attempts to fit the
observed spectra by measuring line EWs and deducing column densities for the different 
ions.  We have searched for the combination of physical components that can be 
produced in nature in 
an environment where low density gas is exposed to a typical AGN continuum. 
These components were then combined in a realistic manner, 
taking into account screening, attenuation of the radiation field, etc.

\subsubsection{Multi-component models}
We have examined the hypothesis that the spectral changes 
observed arise purely as a result of variations in the opacity of the 
absorbing gas which are caused solely by changes in the intensity 
of the ionizing continuum. 
Thus, for this experiment, we  assume 
the shape of the SEDs is the same for both the low- and high-states
and only differ in total luminosity.
 We start by
calculating a variety of models in order to mimic  the 
 low-state spectrum. 
 Each model is made of several emission and absorption components.
The absorbers are specified by  $U_{OX}$, $N$ and the absorption (i.e.
line of sight) covering factor. 
The latter is assumed to be the same for all absorbers (but see also \S4). 
The absorbers are assumed to be aligned such that the observed 
spectrum is the result of the intrinsic continuum passing through them all.
The emission components are specified  by their $U_{OX}$ and $N$ 
(that are not necessarily the same as those of the absorbers), their covering factor 
and whether or not they are occulted  by the absorbing gas.

The dynamics and kinematics of the absorbing gas are important factors in comparing the data
with the model. Following numerous UV observations, and our analysis in Paper I,
we assumed internal motion in the gas which is much larger than the thermal
motion. 
This ``turbulent velocity'' is assumed to have a Doppler profile and the velocity quoted 
is the Doppler $b$ parameter. 
Paper I showed that 
 {\it all absorption lines} with good S/N and sufficient resolution
(the lines of \ion{O}{7}, \ion{Ne}{10}, \ion{Mg}{12} and \ion{Si}{14}) can be characterized by
 two kinematic components.  The central
velocities of these are between  -400 and -600 \kms\ and 
between  -1000 and -1300 \kms, relative to the systemic velocity. A good representation 
of the observations can be obtained with two Doppler profiles each with $b \simeq 250 $ \kms.
 
The relative EW and covering factor of the two velocity components are critical to our modeling of
the source. 
The observations show that the EW ratio of the  velocity components is about 1:0.7 
(lower:higher) in all lines with sufficient S/N (see below).
Since this ratio is seen in several saturated lines, 
it can be interpreted as the result of different absorption covering factors
for each component.
In this case, the lower-velocity component has a covering factor of 0.8--1.0 and the larger-velocity
component a covering factor of 0.6--0.8. 
However, the composite profiles also include unsaturated lines that
seem to have similar shapes. In this case, the different EWs are due to different column densities.
We cannot distinguish between the two possibilities since we do not have 
high-quality, high-resolution profiles for
many weak lines. Given these uncertainties, we investigated two cases where the relative column densities for the two velocity components are in the ratio of 1:0.7. 
In one case the covering factor of all absorbers is unity and in the other case
it is of order 0.8.

In summary, each ionization component of the models presented in this paper is made of
two kinematic components. For the outflow velocities we chose -500 and -1000 \kms, and 
assumed a turbulent velocity of 250 \kms\ for all components.
This implies that the total derived column density of a certain  ionization  component
is {\it the sum} of the column densities in the two kinematic components, even for saturated
lines (since there is very little velocity overlap between the two).
For brevity, in the rest of this paper we just quote the total column densities. 
Given these assumptions, 
we used the EWs measured in Paper I, combined with a few new
measurements, to obtain the various column densities and optical depths.  
The more important lines that were used to constrain our models,
and their adopted column densities, are listed in Table 3.

\begin{deluxetable}{lccccc}
\tablecolumns{6}
\tablewidth{0pt}
\tablecaption{Column Densities\tablenotemark{a}
\label{colden}}
\tablehead{
\colhead{} &
\colhead{} &
\multicolumn{4}{c}{model $\log(U_{\rm ox})$} \\
\colhead{Ion \& Line} &
\colhead{measured\tablenotemark{b}} &
\colhead{$-2.4$} &
\colhead{$-1.2$} &
\colhead{$-0.6$} &
\colhead{total}   }
\startdata
%
%
\ion{Si}{14} $\lambda 6.182$ &  $>17.50$\tablenotemark{c} & \nodata & \nodata & \nodata & \nodata \\
\ion{Si}{14} $\lambda 5.217$ &  $17.94\pm0.10$            & \nodata & \nodata & \nodata & \nodata \\
\ion{Si}{14} $\lambda 4.947$ &  $17.83\pm0.20$            & \nodata & \nodata & \nodata & \nodata \\
\ion{Si}{14} adopted value   &  $17.90\pm0.20$            & 12.50    & 17.00    & 17.48   & 17.60      \\
\ion{Si}{13} $\lambda 6.648$ &  $>16.87$\tablenotemark{c} & \nodata & \nodata & \nodata & \nodata \\
\ion{Si}{13} $\lambda 5.681$ &  $>17.23$\tablenotemark{c} & \nodata & \nodata & \nodata & \nodata \\
\ion{Si}{13} $\lambda 5.405$ &  $17.57\pm0.20$            & \nodata & \nodata & \nodata & \nodata \\
\ion{Si}{13} $\lambda 5.286$ &  $17.71\pm0.25$            & \nodata & \nodata & \nodata & \nodata \\
\ion{Si}{13} adopted value   &  $17.65\pm0.20$            &   14.68 & 17.31   & 16.94   & 17.46   \\
\ion{Si}{12} $\lambda 6.718$ &  $16.10\pm0.06$            &  15.57  & 16.55   & 15.41   & 16.62    \\
\ion{Si}{11} $\lambda 6.778$ &  $16.29\pm0.06$            &  16.43  & 16.15   & 14.11   & 16.61    \\
\ion{Si}{10} $\lambda 6.859$ &  $16.83\pm0.05$            &  16.90  & 15.88   & 13.10   & 16.94    \\
\ion{Si}{9} $\lambda  6.931$ &  $16.66\pm0.06$            &  17.04  & 15.27   & 11.68   & 17.05    \\
\ion{Si}{8} $\lambda  6.999$ &  $16.66\pm0.08$            &  16.81  & 14.26   & 9.93    & 16.81    \\
\ion{Si}{7} $\lambda  7.063$ &  $16.26\pm0.20$            &  16.13  & 12.71   & 7.80    & 16.13    \\
\ion{S}{16} $\lambda  4.729$ &  $>17.29$\tablenotemark{c} & \nodata & \nodata & \nodata & \nodata  \\
\ion{S}{16} $\lambda  3.991$ &  $17.58\pm0.13$            & \nodata & \nodata & \nodata & \nodata \\
\ion{S}{16} adopted value    &  $17.58\pm0.13$            & 10.13   & 16.23   & 17.16   & 17.21   \\
\ion{S}{15} $\lambda  5.039$ &  $16.87\pm0.09$            & \nodata & \nodata & \nodata & \nodata \\
\ion{S}{15} $\lambda  4.299$ &  $17.08\pm0.25$            & \nodata & \nodata & \nodata & \nodata \\
\ion{S}{15} adopted value    &  $17.00\pm0.20$            & 12.70   & 16.90   & 17.00   & 17.25   \\
\ion{Mg}{8} $\lambda  9.506$ &  $16.58\pm0.05$            & 16.77   & 14.46   & 11.13   & 16.77   \\
\ion{Mg}{9} $\lambda  9.378$ &  $16.39\pm0.06$            & 17.06   & 15.13   & 12.61   & 17.07   \\
\ion{O}{8} $\lambda  14.821$ &  $18.67\pm0.25$            & \nodata & \nodata & \nodata & \nodata \\
\ion{O}{8} $\lambda  14.634$ &  $18.59\pm0.14$            & \nodata & \nodata & \nodata & \nodata \\
\ion{O}{8} adopted value     &  $18.63\pm0.25$            & 17.65   & 18.05   &  17.55  & 18.28  \\
\ion{O}{7} $\lambda  17.396$ &  $18.03\pm0.20$            & \nodata & \nodata & \nodata & \nodata \\
\ion{O}{7} $\lambda  17.200$ &  $18.04\pm0.20$            & \nodata & \nodata & \nodata & \nodata \\
\ion{O}{7} adopted value     &  $18.04\pm0.20$            &  18.39  & 16.84   & 15.49   & 18.40   \\
\ion{O}{6} $\lambda  21.01 $ &  $>16.42$\tablenotemark{c} & \nodata & \nodata & \nodata & \nodata \\
\ion{O}{6} $\lambda  19.341$ &  $>17.00$\tablenotemark{c} & \nodata & \nodata & \nodata & \nodata \\
\ion{O}{6} adopted value     &  $>17.00$            &  17.96  & 14.34   &  12.23  & 17.96    \\
\ion{O}{5} $\lambda  22.334$ &  $>16.34$\tablenotemark{c} & \nodata & \nodata & \nodata & \nodata \\
\ion{O}{5} $\lambda  19.924$ &  $17.07\pm0.40$            & \nodata & \nodata & \nodata & \nodata \\
\ion{O}{5} adopted value     &  $17.07\pm0.40$            &  17.54  & 12.59   &  9.3    & 17.54    \\
\enddata
\tablenotetext{a}{Log of column density in units of cm$^{-2}$.}
\tablenotetext{b}{Column densities and uncertainties  derived from the EWs in Paper I.
The EW was divided in the ratio 1:0.7, then the column
density corresponding to the 1/1.7 part was computed assuming a Doppler
width of 250 \kms . The derived column density was multiplied by 1.7 to
result with the total column density listed here.  Uncertainties were
calculated from the EWs uncertanties. See text for details.}
\tablenotetext{c}{Saturated line, lower limit only.}
\end{deluxetable}

\subsubsection{The silicon and sulphur line method}
A major clue for the conditions in the absorbing gas is obtained from EW measurements 
 of various lines in the 5--7.1~\AA\ band. 
 This wavelength range contains lines from \ion{Si}{7} to
\ion{Si}{14} as well as the strongest lines of \ion{S}{15} and \ion{S}{16}. 
Most of these lines are not blended.
The range of ionization and excitation is very large and represents the ionization of almost
all line-producing ions in the \chandra\ spectrum.
The column densities deduced from these EWs, assuming the two component profiles,
are given in Tables 1 \& 3. The atomic data for the lines
are known either from standard calculations (f-values for the H-like and He-like transitions) or
from the recent work of Behar and Netzer (2002) discussing the inner-shell
lines of silicon, sulphur and other elements.

We have developed a simple algorithm to compare the measured optical depths of all 
the observed silicon and sulphur lines  with the results of the
photoionization calculations. We first compute a large grid of
models with calculated optical depths for a large
range of ionization parameters and column densities. We then pick up to
four models at a time and compare the {\it combined} optical depth for
the chosen set of lines with the  values deduced from the observations for the
larger column density (lower outflow velocity) kinematic component. 
The best combination of models is obtained by
minimizing the differences between the observed and the calculated optical depths
in all 10 lines (the 9 silicon and sulphur lines listed in Table 1 plus a line of \ion{Si}{7}).  
The result is a list of up to four models, with 
various  column density and $U_{OX}$,
whose combination gives the best match to the observed column densities. 
Using this procedure we find that at least three ionization components (i.e. six kinematic 
components) are required to fit the data to within the observational accuracy. 
The reason is the very large difference in ionization between the lowest
(e.g. \ion{Si}{7} and \ion{Si}{8}) and the highest (e.g. \ion{S}{16}) ionization lines.
This requires at least one component to fit the EWs of the lowest ionization lines, one
to fit the intermediate ionization lines and one for the highest ionization lines.

Experimenting with various modifications of the method suggests that
there are several combinations of
three or four ionization components that give similar quality fits to the
observed optical depths of the silicon and sulphur lines. 
All combinations share the same general properties ---
they all require at least three very  different sets of physical conditions (i.e. ionization parameters)
and three similar column densities. A generic model with the required
properties is the following three-component model 
(all column densities refer to the total column of the two kinematic components):
a low ionization component
with log$(U_{OX})=-2.4 \pm 0.1$ and log$(N)=21.9 \pm 0.1$, 
a medium ionization component
with log$(U_{OX})=-1.2 \pm 0.2$ and log$(N)=22.0 \pm 0.15$,
and a  high ionization component
with log$(U_{OX})=-0.6 \pm 0.2$ and log$(N)=22.3 \pm 0.2$.
The uncertainties on $U_{OX}$ and $N$ in the case of the low ionization component
arise primarily from the uncertainty on the slope of the ionizing continuum.
Relaxing this constraint allows a somewhat smaller column density (see \S3.2.4).
The larger uncertainties on the parameters of the higher ionization  components are due to the
fact that gas with such properties produce similar EWs for medium
(e.g. \ion{Si}{11}) and high (e.g. \ion{Si}{14}) ionization lines over a larger range of 
ionization parameters and column densities.
The three theoretical ionization components are plotted over a large wavelength range in Fig. 5,
and a more detailed view of the 4.5--7~\AA\ range is shown in Fig. 6.

\begin{figure}
\centerline{\includegraphics[width=10cm]{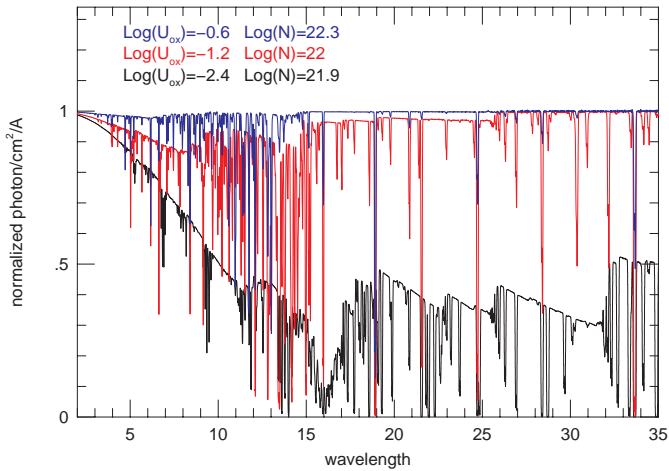}}
\figcaption
{The three generic components that were used to model the low-state spectrum of NGC\,3783.
All components are shown on the same normalized scale where 1 is the
incident continuum level.
\label{fig5_label}
 }
\end{figure}
\begin{figure}
\centerline{\includegraphics[width=10cm]{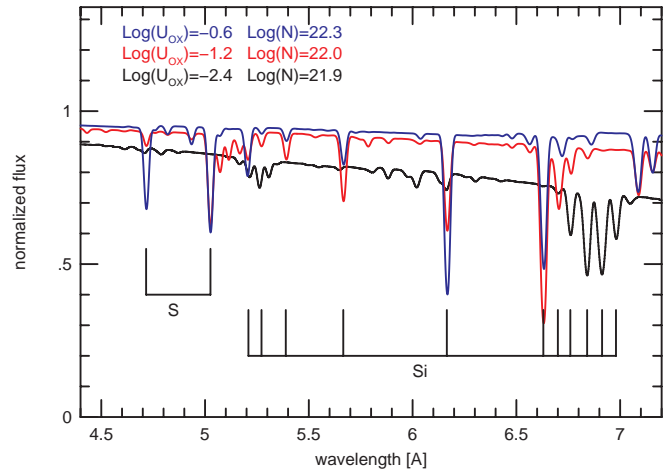}}
\figcaption
{A blown up version of Fig. 5 showing the 4.5--7~\AA\ region used to constrain the model
parameters (see text). Note that all three models are required to explain the large
range of ionization (from \ion{Si}{7} to \ion{S}{16}.)
\label{fig6_label}
 }
\end{figure}

\subsubsection{Intrinsic continuum and global model properties}
The next step is to test the combination of components that fit the optical depths of the
5--7.1~\AA\ silicon and sulphur ions best 
over other wavelength bands. This requires  
 the shape of the  underlying ionizing continuum to be defined more
precisely.
The slope of the short wavelength continuum can be measured directly 
if  bound-free absorption is negligible.
However, Fig. 5 clearly shows that two of the generic absorbers are characterized
by  large opacity 
at wavelengths as short as 4--5~\AA. Thus a unique determination of the 2--5~\AA\
continuum slope (the part available for the grating observations) is not
trivial. More specifically, the exact \ion{O}{7} column density
in the lowest-ionization component strongly influences the short wavelength continuum
shape, and the uncertainty on this column gives rises to an uncertainty on the slope.

Given the above constraints on the model properties, and the observed 2--5~\AA\ continuum,
we have experimented with various power-law continua with the requirement
that both the short
wavelength continuum and the measured EWs of the silicon and sulphur lines, are within
the observational uncertainties. All fits were performed on the low-state spectrum.
We found  that a single power-law 
(where the number of photons per unit energy, $E$, is proportional 
to $E^{-\Gamma}$) with $\Gamma=1.65 \pm 0.15$ is a good
overall representation of the low-state spectrum. 
We note that the limited data quality cannot
exclude the possibility of some steepening of the intrinsic continuum at long
 ($\lambda >20$\AA) wavelengths. 

We note in passing the different continuum slope adopted by K03 in their recent
modeling of the source. These authors 
did not consider the two spectral states found here and fitted the full
\chandra\ data set.  As a result, the SED they adopted 
is based on the combined (high- and low- state) spectrum, and hence is different from
ours. This is also the reason why their powerlaw slope is flatter than
the one adopted by Kaspi et al. (2001), who fitted the high state only.
K03 also include in their SED a low-energy blackbody component.

Another uncertainty on the intrinsic
SED arises from the fact that some of the high energy photons are likely to be
produced far away from the source, due to Compton scattering and reflection of the
central continuum radiation. This can be very noticeable in faint
sources, like NGC\,3516 during 2000 (see Netzer et al.
2002). Scattering by gas in a low state of ionization is 
strongly wavelength dependent, and for NGC\,3783
can amount to a change of $\sim 0.1$ in the derived value of $\Gamma$.
As a result, we might have overestimated the
column density of the lowest ionization component (the only component with significant 
influence on
the slope of the short wavelength continuum) which  would results in overestimating the 
column densities of low ionization species such as \ion{O}{5} and \ion{O}{6}. 
\S4.1 contains some discussion about the lowering of the column density in our low $U_{OX}$
component and a more detailed examination of this effect is
deferred to a separate paper (George et al. 2003, in preparation).

Finally, a broad, relativistic iron K$\alpha$ line can also be important
in affecting the short wavelength continuum slope. The presence or absence of such a feature is
an open question. Kaspi et al. (2001) see no evidence
for this feature in the 2000 \chandra\ observation. Given the uncertainties discussed above,
and the large amount of absorption at almost all wavelengths, we cannot resolve this issue even
with our much improved data.

Given the chosen continuum and the three ionization components, we have calculated 
a combined multi-component model that covers the entire wavelength range. 
As stated above, the model 
includes two kinematic components for every ionization component.
The theoretical
spectrum is compared with the observed low-state spectrum in Fig. 7.
The results are very good. In fact, we could not find another combination of models
(three or more absorbers) that is significantly superior.
Not only does the model provide a good overall fit to many lines of 
all elements, but it also predicts the strengths of the 
deep bound-free edges correctly.
Thus, the constraints on the 5--7.1~\AA\ silicon and sulphur
lines are enough to completely specify a model for the entire spectrum.
\begin{figure*}
\centerline{\includegraphics[width=17.1cm]{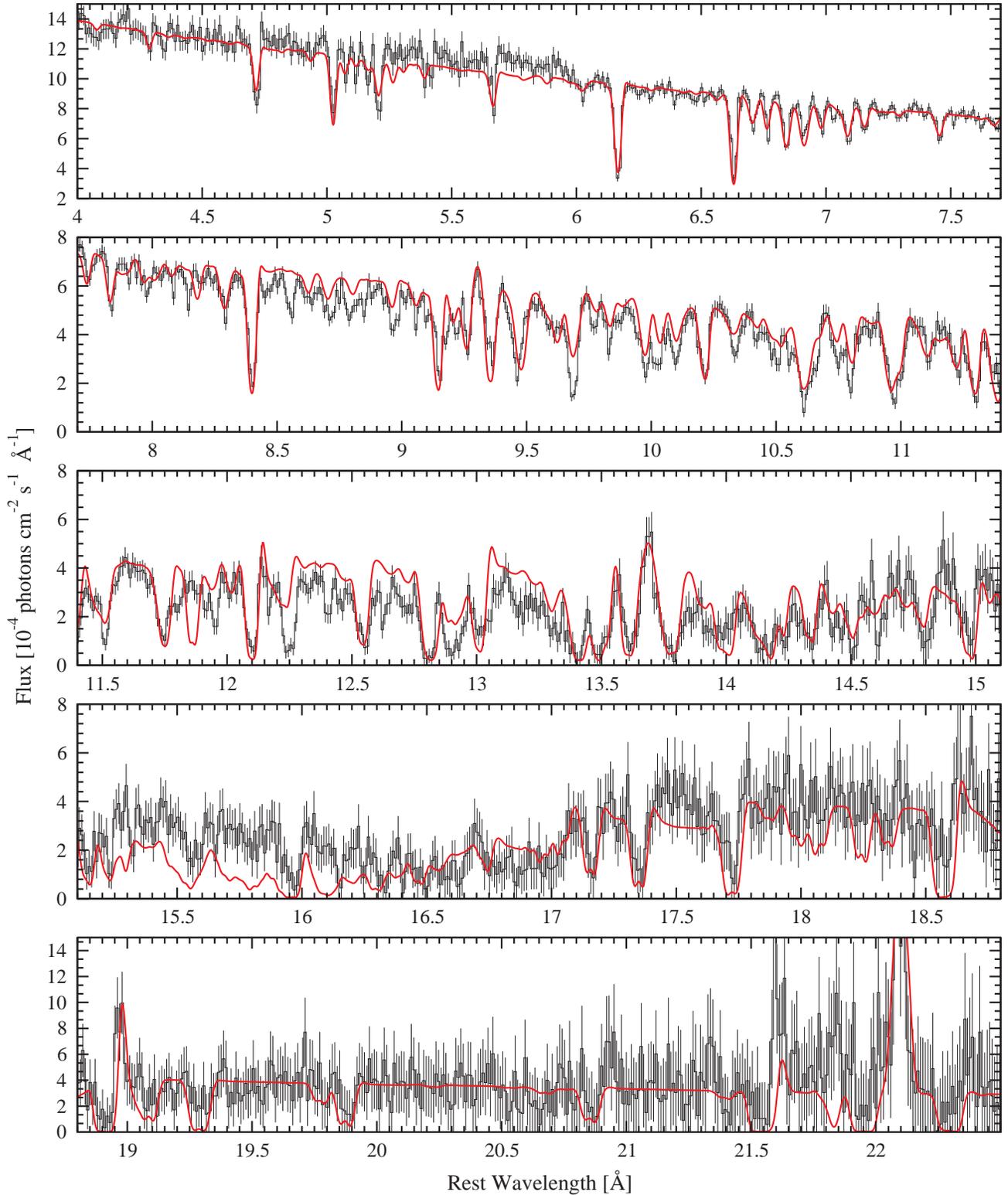}}
\figcaption
{A comparison of the low state spectrum with the six component model (three ionization
components each split into two kinematic components).
Note that most of the apparent discrepancy at around 5--6~\AA\ is due to
a known calibration problem (H. Marshall, private communication).
\label{fig7_label}
 }
\end{figure*}
\begin{figure*}
\centerline{\includegraphics[width=17.5cm]{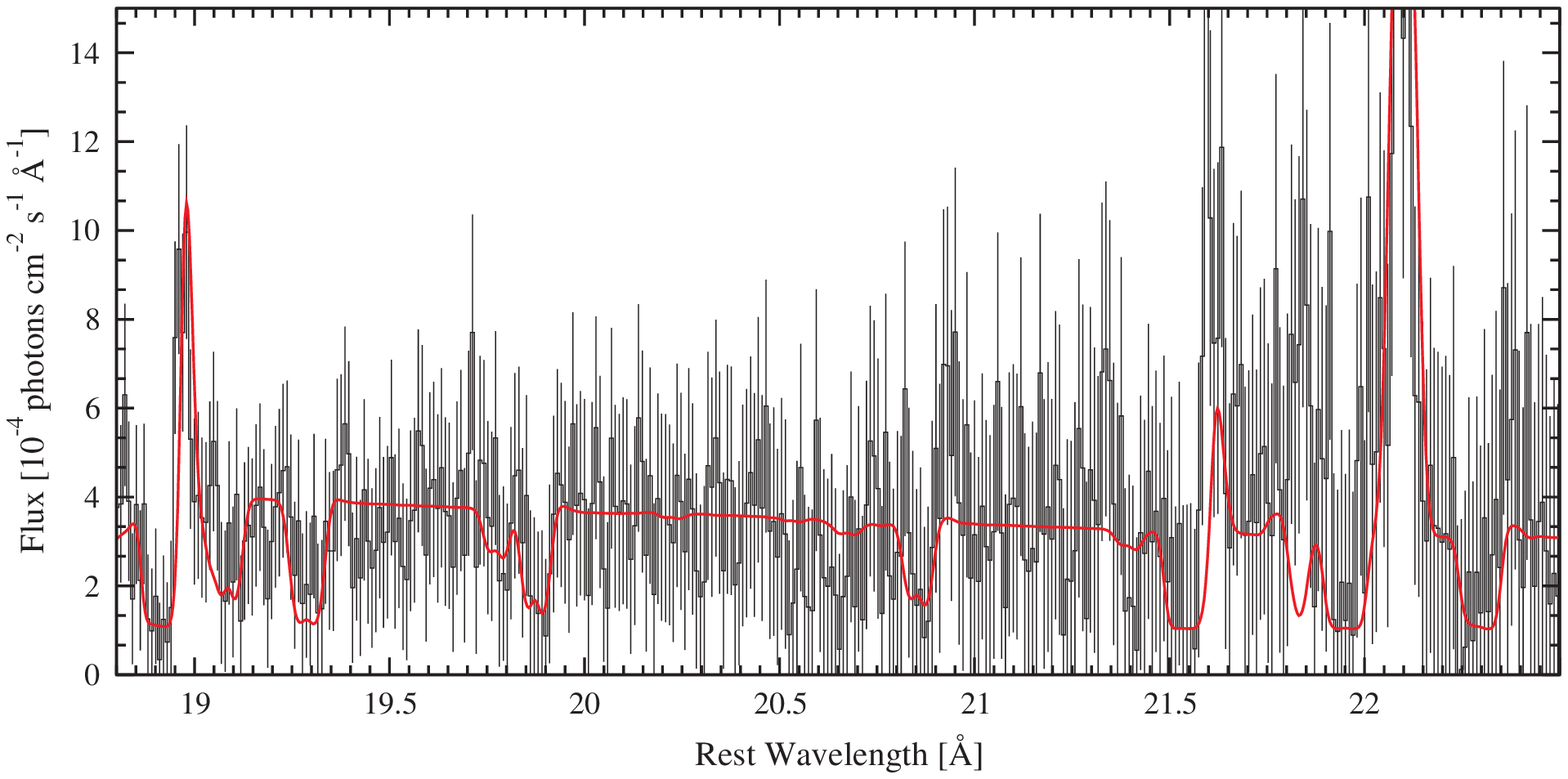}}
\figcaption
{Model vs. observation for the long wavelength spectrum assuming $\Gamma=1.5$ and
 a line-of-sight covering factor of 0.85 for all absorbers.
\label{fig8_label}
 }
\end{figure*}

\subsubsection{Deficiencies in the Modeling}

As demonstrated above,
the low-state absorption spectrum of NGC\,3783 can be satisfactorily explained by 
our combination of line-of-sight absorbers.
The agreement with the observations is very good for all lines with
 $\lambda<15$\AA.  However, there is a noticeable discrepancy at
around 16\AA.
 At this wavelength range there are two main sources
 of opacity -- the \ion{O}{7} bound-free
continuum (edge at 16.7\AA), and the iron M-shell UTA feature (Behar, Sako \& Kahn 2001).
Our photoionization calculations suggest
a disagreement between the two in the sense that the strongest iron lines 
observed are due to ions indicative of a lower level of ionization than that
indicated by the strongest oxygen lines observed. 
Specifically, the peak UTA absorption corresponds to \ion{Fe}{8}--\ion{Fe}{10}.
For such a level of ionization, our models predict 
most of the oxygen being  in the form of \ion{O}{3}--\ion{O}{6}.
However, the observations indicate that most of the oxygen
bound-free absorption is due to \ion{O}{7}.
The possible origin of this discrepancy  is discussed in
\S4.                                                      

Regarding the spectrum beyond  $19$\AA, the comparison between the model and the 
observations is limited by the poorer S/N.
While the model shown in Fig. 7 is consistent with the data, 
the predicted depths of several absorption lines in this range
(in particular those due to \ion{O}{5} and \ion{O}{6}) 
seem to be stronger than actually observed. 
This may be a real shortcoming of the model (see comments in \S3.2.4 and in \S4). 
It may also be the result of the assumed absorption
covering factor. As already mentioned, line-profile analysis suggest a mean (over
the line profile) covering factor of 0.7--1.0, and all models considered so far assumed the extreme
case of  an absorption covering factor of unity. 
 The ``leakage'' of the incident continuum makes
little difference at short wavelengths except for a need to 
somewhat decrease the assumed $\Gamma$.
However, leakage 
 can influence the comparison at long wavelengths much more.
This is illustrated in Fig. 8 which shows a model with $\Gamma=1.5$ and a line-of-sight covering factor of 0.85.
Indeed, the agreement at long wavelengths is much better. The differences between the two assumed slopes and
covering factors are well within the model uncertainties.

We have attempted to model the high-state spectrum
by  assuming the same ionization components and by changing the luminosity and the ionization parameter
by the factor inferred from the observed increased luminosity in the short
wavelength continuum ($\sim 1.5$). The result is a very poor fit to the long wavelength
 continuum  luminosity and
to the intensities of many lines.  The reason is that such a small luminosity
 change results in an  opacity change which is too small to account for  the very
large difference (factor $\sim 4$) between the low and the high state spectra  at long 
wavelengths. This means that the high and the low state spectra are
inconsistent with the assumption of a simple response of the gas to continuum luminosity variations.

There is a simple and satisfactory solution for the
high-state spectrum that involves the appearance of an additional continuum component.
This second, long wavelength component (sometimes referred to as a ``soft excess'')
appears only during  high state. We have therefore modeled
the high-state continuum by two powerlaws: the low-state
power law ($\Gamma=1.6$)  and a steeper component with $\Gamma=3.2$
which dominates at long wavelengths.
We find a satisfactory solution for the high-state spectrum when 
the $\Gamma=3.2$  component
emits 40\% of the flux of the $\Gamma=1.6$ component at 12.398~\AA\ (1 keV). 
(We note that this is not a unique combination, and fits of similar quality can be 
obtained for other combinations of slope and normalization.)
The two continua are shown in Fig. 3, along with 
 the high and the low state spectra.
 Given the two component continuum, we find a very good agreement for the
high-state line and continuum spectrum by assuming {\it no opacity variations}.
The quality of the fit is similar to the low-state fits shown in Figs. 7 \& 8.

 To further illustrate this point in Fig. 9 we show the high-state spectrum 
divided by the low-state spectrum (binned to reduce the noise)
along with the ratio of the two assumed continua. The diagram shows that
any remaining spectral features are entirely consistent with the noise.
The S/N in this diagram is not high enough to completely rule out some
opacity-like variations at  long wavelengths. In particular, there is some
excess emission near the \ion{O}{7}  and \ion{O}{8} recombination edges that
may hint to extra emission in the high-state spectrum (note that we do not
expect the emission features to disappear by this division). 
However, such opacity
variations must contribute very little to the spectral variations 
observed at long wavelengths.

\begin{figure}
\centerline{\includegraphics[width=10cm]{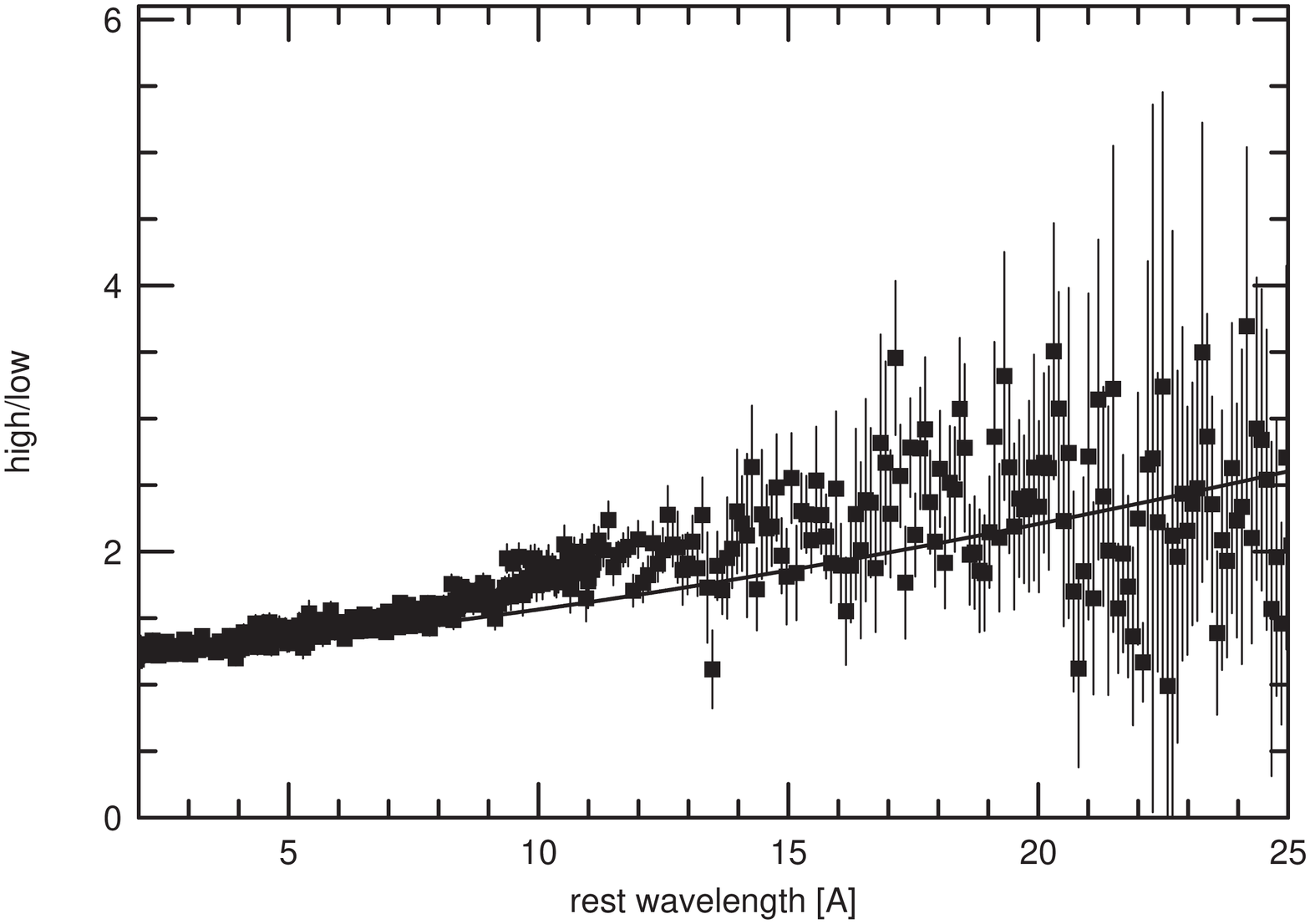}}
\figcaption
{A ratio spectrum of NGC\,3783: Points with error bars are the result of
dividing the high-state spectrum by the low-state spectrum. The binning is
0.04\AA\ for $\lambda<11$\AA\ and 0.1\AA\ at longer wavelengths.
The smooth line is the same division
applied to the two continua shown in Fig. 3.
\label{fig9_label}
 }
\end{figure}

Finally, the emission-line spectrum of NGC\,3783 can be explained by assuming 
X-ray emitters with the same properties found for the ionized absorbers. The 
emitted line photons are probably observed through the absorber, as
indicated by the combined emission-absorption profiles of several resonance lines.
The model is problematic at around 21.8\AA, where it fails to reproduce
the \ion{O}{7} intercombination emission line.
The emission covering factor required is about 0.1 (see \S4.1) consistent with the value
of approximately 0.05--0.15 obtained by Behar et al. (2003).

\section{Discussion}

This section discusses the main results of the new analysis of all the \chandra\
observations of NGC\,3783 performed to date.
The reader is referred also to an appendix where we give a 
detailed comparison with the recent work of Krongold et al. (2003).

The main new findings of our work are the large increase in flux between the
low-  and high-states, which is {\it not} accompanied by any significant 
changes in the properties of the absorption. 
The central source in NGC\,3783 
fluctuates between two such flux states on a 20--120 days time scale,
 and the best description of the
observed variations is the appearance and disappearance of a ``soft
excess'', low energy continuum component. These lead to important conclusions 
regarding the nature of the continuum source.
Other important conclusions are the multi-component nature of the absorber and
the large range of ionization. Below we discuss some interpretations.

\subsection{Physical properties of the absorbing gas}
We found that three ionization components with  different
properties and a large range of ionization, are required to fit the
spectrum of NGC\,3783. Each of those
components is split into two kinematic components, as implied by
the observed line profiles reported in Paper~1.
 The three ``generic absorbers'' of our model are listed in Table 4 and shown in Figs. 5 \& 6.
As explained above, the number of real components and the exact values of  $U_{OX}$ and $N$
are not very meaningful since several somewhat different combinations of ionization components
 give equally good fits. Rather, the values chosen here are representative, "average" 
ionization parameters, and the total column densities required.
Our modeling of the low-state spectrum assumes the underlying X-ray continuum is 
a single power-law. However, the S/N of the long wavelength spectrum is not high enough to
completely rule out the presence of an 
additional softer continuum component, similar to the one seen during the high states but of much
lower luminosity. We also note that there is little (if any) response of the absorbing gas to the
short ($\sim 2$ days) time scale continuum variations.  Thus our measurements 
are models represent some time-averaged spectral properties.

\begin{deluxetable}{ccc}
\tablecolumns{3}
\tablewidth{0pt}
\tabletypesize{\footnotesize}
\tablecaption{Parameters of the Three  Generic Models
\label{Model}}
\tablehead{
\colhead{Model } &
\colhead{log$(U_{OX})$ } &
\colhead{log($N$)}
}
\startdata
Low ionization  &   -2.4        &  21.9          \\
Intermidiate ionization  &    -1.2    &  22.0            \\
High ionization &   -0.6      &  22.3     \\
\enddata
\end{deluxetable}

The main difference between the present model and the one given by Kaspi et al. (2001) is
the presence of the lowest ionization component. 
Only the first \chandra\ observation (obtained in 2000 January) was available to Kaspi et al., 
and hence was insufficient to reveal the two spectral states. 
Furthermore, this relatively short observation resulted in the S/N  at $\lambda > 18$\AA\ which
was insufficient to reveal the true shape of the continuum.
Hence Kaspi et al. naturally (but incorrectly) assumed the underlying continuum was
a single power-law for  this high-state spectrum. 
As a result, their preferred model contained 
a component with a strong \ion{O}{7} bound-free edge, but very little  absorption at
longer wavelengths (i.e. at $\lambda > 18$\AA\ the underlying continuum was "recovered").
Our new observations do not show such a recovery. In fact, the absorption in the long wavelength
region is dominated by features of K-shell carbon and nitrogen and L-shell nitrogen and oxygen.
 
The presence of a lower ionization component is also a key point in the K03 work. In their
study they assume two absorbers with log$(U_{OX})=-2.77$ (i.e. a factor of about two lower than
our lowest ionization component) and log$(U_{OX})=-1.23$ (very similar to our imtermediate
ionization component).  Their fit  does not require  a third, very ionized component. 
Our reason for suggesting this higest-ionization  (log$(U_{OX})=-0.6$)
 component is based on the fit to the lines
of \ion{Si}{14}, \ion{S}{15} and \ion{S}{16}. When this component is not included, the discrepancy
in column densities of these ions is a factor 2--3. Several of the lines of these ions
are in fact missing from the K03 analysis. Moreover, some of the highest -ionization
lines are badly saturated, (e.g. \ion{Si}{14}$\lambda 6.18$\AA). Thus a  comparison
of the observed and the calculated EWs of those lines, as performed by K03,
is not sufficient for evaluating the ionic column densities.

Perhaps the largest observed discrepancy between model and observation is due
to absorption by the iron M-shell UTA. This feature has been observed
in several other AGN (Steenbrugge et al. 2003 and references therein) with similar 
shape and central wavelength. 
Its smooth  shape  is due to the large number of absorption
lines of many iron ions and the wavelength of largest absorption is ionization
dependent. Our modeling of this spectral region includes all the lines in the Behar et al. (2001)
calculations (i.e. not only the shorter published list of lines).
 The central observed wavelength of the feature in
NGC\,3783 is at around 16.4\AA\ suggesting that most of the contribution is
due to \ion{Fe}{8}--\ion{Fe}{10}. For the assumed  SED, this corresponds to gas
whose dominant oxygen ions are  \ion{O}{3}--\ion{O}{6}.
In our modeling, most  long
wavelength absorption is due to the component with log$(U_{OX})=-2.4$ (Fig. 5). 
The dominant ions in
this components are \ion{O}{6}--\ion{O}{7} and   
\ion{Fe}{10}--\ion{Fe}{12}. This corresponds to peak UTA opacity at around 16.1\AA.
As shown in Fig. 7, the model fits the observed oxygen lines and continua reasonably well  
but badly misses the position of the UTA absorption.

We have experimented with various other models to try to eliminate this
discrepancy. For example, an overabundance of iron relative to
oxygen may decrease  the level of ionization of iron due to the increased iron
opacity. Experimenting with iron abundance which is three times larger
resulted with changes that are not large enough to explain the observed
discrepancy. We have also experimented with models that include several low ionization
components, instead of the 
single low $U_{OX}$ component shown here. This, again, gave very little improvement.
 It seems that the apparent conflict between the oxygen and iron
ionization cannot be resolved by these models and we suggest
 two alternative explanations.
\begin{enumerate}
\item
The absorbing gas in NGC\,3783 may not be in ionization equilibrium due to
its low density and the rapid flux variations on time scales that are 
shorter than the recombination and ionization time scales. This is probably not very important
for the short wavelength flux since its variability time scale is short and the amplitude not very
large. The gas responding to this continuum is probably at some
 mean level of ionization. This however is not the case 
for the soft excess continuum which varies on a much longer time scale with a much larger amplitude.
This can contribute, significantly, 
to the ionization of the lower ionization species. Different ions react on different time scales
and the gas may never reach an equilibrium. Since all our models 
assume steady state gas, they may not be adequate to describe the absorbing gas properties. 
In particular, the iron and oxygen recombination times may be different enough to result in
gas where iron is less ionized or oxygen is more ionized compared to the equilibrium situation. 
The complicated issue of time dependent ionization is beyond the scope of this paper. 
\item 
The ionization balance of iron depends critically on 
low temperature dielectronic recombination (DR) rates that are not well known for the 
iron M-shell ions.
Experience with L-shell iron ions, whose DR rates have been measured and
computed recently (Savin et al. 2002 and references therein), shows that in almost all cases the
previous low temperature DR rates were consistently smaller than the newly
calculated values. Assuming a similar effect in the M-shell ions, we can
envisage a situation where, in the absence of realistic DR rate, the entire
ionization balance of iron is shifted toward  higher ionization. 
 Thus, more realistic low temperature DR rates may bring models to better
agreement with the observations. We note that low temperature DR rates are also not available
for the magnesium and silicon ions that we have modeled. For some of these ions (\ion{Mg}{8} and \ion{Mg}{9})
these are probably not very important and for others (the relevant silicon ions) 
the situation is less clear (D. Savin, private communication but see also Gu 2003).
Nevertheless, since we do obtain  good agreement between the calculated EWs of these lines 
and the EWs of other lines observed in the spectrum, we 
did not consider changing those rates in the calculations.
\end{enumerate}

It is interesting to  note that the K03 model assumes a
lower level of ionization for all elements, including iron, and hence provides a better fit
to the UTA.  However, their lower ionization component fails to fit the
\ion{Si}{10} and \ion{Si}{11} lines around 6.8\AA. Our model was designed
to give a very good fit over the 5--7\AA\  wavelength range and hence produces
a lower quality fit to the UTA feature. Thus, there seem to be no satisfactory 
solution based on a single low-ionization component that explains all these features.

An interesting and somewhat problematic issue is the abundance of \ion{O}{6},
 which can also be clearly
observed in the far UV (Gabel et al. 2003a; Gabel et al., 2003b).
Paper I found even the K$\alpha$ line of
\ion{O}{6} undetectable with the current S/N. However, revisiting this
measurement more carefully, and with more accurate atomic data,
we can now detect several \ion{O}{6} lines. From two such lines,
both of which are probably saturated, we have 
obtained a lower limit of $10^{17.0}$ cm$^{-2}$ on the column density of \ion{O}{6}. 
This is much higher than the value ($10^{16.0 \pm 0.3}$ cm$^{-2}$) obtained 
from the \xmm\ RGS spectrum (Behar et al. 2003). 
Moreover, the values predicted from our model ($10^{17.96}$ cm$^{-2}$) is even larger.
Such a large value is problematic for several reasons. It may be in contradiction with
the UV observations of the source (see below), it is much larger than the values obtained
from the RGS observations, and it results in a prediction of a strong absorption
line at around 21.87~\AA\ which is not seen in the spectrum (Fig. 7, the absorption line underneath
the \ion{O}{7} intercombination line).

An explanation of the discrepancy between the values for the \ion{O}{6}
column density measured by \chandra\ and \xmm\  is the crowded region of the
spectrum where \ion{O}{6} K$\alpha$ resides. In particular, the lower spectral
resolution of the \xmm\ RGS makes it extremely difficult to resolve the weak
\ion{O}{6} absorption line at 22.01~\AA\ in the presence of the bright
\ion{O}{7} forbidden emission line (22.10~\AA). We also wish to note that
with the lack of laboratory measurements for the \ion{O}{6} wavelengths,
high-resolution astrophysical spectra such as the present one constitute the
best determination of these wavelengths currently available.  The \chandra\ data,
along with the assumption that \ion{O}{6} is outflowing at the same velocity
as \ion{O}{7}, places the \ion{O}{6} K$\alpha$ line at a rest-wavelength of 22.01~$\pm$~0.01~\AA.
This value is in agreement with our model
wavelength of 22.005~\AA, calculated with the HULLAC atomic code, as well as with
the value measured in NGC~5548 (Steenbrugge et al., 2003), but is somewhat different from the value of
22.05~\AA\ calculated by Pradhan (2000) using the R-matrix method. For more information see Behar 
and Kahn (2002).

A \ion{O}{6} column density less than $10^{17}$ cm$^{-2}$ 
 is also inconsistent with
theoretical predictions based on modelling the abundances of
\ion{Mg}{8}, \ion{Si}{8} and \ion{O}{6}. This is illustrated in Fig. 10 where we
show calculations for the  fractional abundance of oxygen, silicon and magnesium
ions as a function of $U_{OX}$ assuming an incident continuum similar to the one used here.
The diagram shows the very similar fractional abundance of \ion{O}{6}, \ion{Mg}{8} and
\ion{Si}{8} as a function of ionization parameter. Using the measured EW of the
 \ion{Si}{8}$\lambda \, 6.998$ line, which is an unblended, easy-to-measure feature, we
can use Fig. 10, and the assumed silicon-to-oxygen abundance ratio, to derive an estimate
for the column density of \ion{O}{6}. This number is very close to the model prediction.
The situation is more complicated regarding magnesium since
several of the relevant lines (around 9.4\AA) are blended with neon lines. However, we
could obtain an estimate of the EW of the \ion{Mg}{8}\,$\lambda 9.506,9.378$ lines that 
are good enough to constrain the \ion{Mg}{8} column density. The \ion{O}{6} column density
based on this measurement is, again, much larger than $10^{17}$ cm$^{-2}$
and in good agreement with the model prediction.
\begin{figure}
\centerline{\includegraphics[width=10cm]{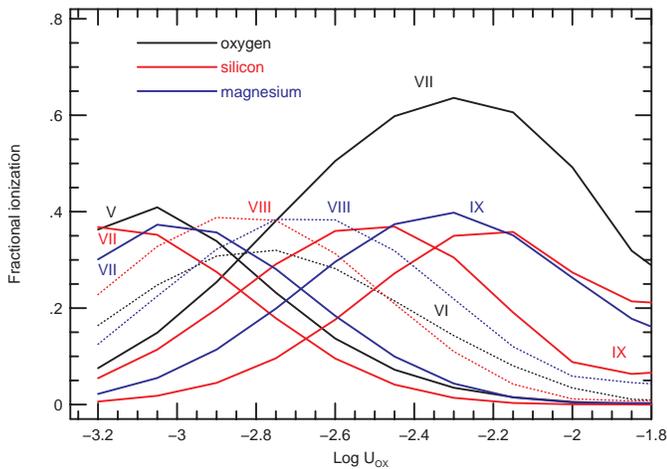}}
\figcaption {
The fractional ionization of oxygen, magnesium
and silicon ions in a low density gas of our chosen composition exposed to the NGC\,3783 low-state
continuum. The curves show the mean ionization of various ions averaged over a column density
of $10^{21.5}$ cm$^{-2}$. Note the great similarity in the ionization of  \ion{O}{6},
\ion{Mg}{8} and \ion{Si}{8} (all plotted with dotted lines).
\label{fig10_label} }
\end{figure}

A comparison with the K03 work shows 
a disagreement by a factor of two in the total (hydrogen) column density of the
lowest-ionization absorber.
One reason for the difference is that K03 analyse the combined (high- and low-state)
spectrum, which is then interpreted in terms of a different amount of bound-free absorption.
Another reason is the somewhat larger oxygen abundance assumed by K03
(the oxygen column densities of the low-ionization components
in the two papers actually differ only by a factor 1.35).
We also note that the inner-shell \ion{O}{6} lines that we have used to determine the
limit on the column density of this ion are missing from the K03 model.
This makes a noticeable difference in the  spectral fitting  at  around 21\AA.

Finally, we have looked for the lowest \ion{O}{6} column density allowed by the observations if
we remove the constraint imposed by the short wavelength continuum slope (\S3.2.4) and keep the ionization parameter unchanged.
This means finding the lowest column density  for the 
log($U_{OX}$)=-2.4 component which is still
consistent with the measured EWs of the 5--7.1~\AA\ silicon and sulphur lines. 
We found this limit to be log($N) \gtrsim 21.7$ for the low-state spectrum, and
log($N) \gtrsim 21.6$ for the high-state spectrum. These lower limits on the 
column densities are consistent with the uncertainties given in Table 1.
Under these assumptions, the column density of \ion{O}{6} in the low-state spectrum
 can be as low as $10^{17.76}$ cm$^{-2}$.
This column is large enough to produce saturated \ion{O}{6} lines at 21.01 and 21.87~\AA.
It also requires a much harder continuum ($\Gamma=1.4$).

\subsection{Outflow velocity and covering factor}

Some of the parameters  our model are based on measured profiles of several
absorption lines that were shown in Paper I to include at
least two components with different ouflow velocities.
The measurements in Paper I also show that the
relative EWs of the two  components
seems to be ionization independent and that most of the lines used for the profile
analysis were heavily saturated.
This suggests that the apparent optical depths of those lines are
determined by the covering factor rather than the line opacity, and 
hence similar to the case seen in  many UV absorption systems in AGN 
(e.g. Barlow, Hamann \& Sargent 1997;Arav et al. 1999).
The covering factor of the lower-velocity components of the \ion{O}{7}, \ion{Ne}{10}
and \ion{Si}{14} lines is in the range 0.8--1.0.  The covering factor of
the larger-velocity component of the same lines, assuming saturated
profiles, is 0.6--0.8.
We also note that the large number of iron lines near the UTA center gives an
independent estimate of the covering factor which is at least 0.85.

Fig. 8 shows that models with a covering factor less than unity provide 
fits to the data of similar quality. 
Using the silicon and sulphur lines (as in \S3.2.3), we find
        three ionization components are again required, each with
        parameters similar to those listed in Table 4. The only difference is
the need to assume a somewhat harder ionizing continuum. The fit to the long wavelength lines
is in fact somewhat better than that produced by the full covering models.
However given the low S/N at those wavelengths, this is not a strong conclusion.
Thus, the data-model comparison cannot constrain covering factor beyond the 
actual observations.
The most important conclusion is that the covering factor is similar 
in low (e.g. \ion{O}{7} and the iron UTA) and in  high
 (e.g. \ion{Si}{14}, \ion{Ne}{10}) ionization lines.

\subsection{Density, location and thermal stability of the absorbing gas}
 
Our measurements and analysis are consistent with no variations in the 
absorbers' properties on time scales of
1--4 days. There are also indications for no spectral changes 
over much longer periods, perhaps a few months.
For instance we find no evidence for any significant, 
narrow ($<$ 1~\AA) spectral features in the high- to low-state
spectral ratio shown in Fig.9.
The reason we cannot constrain the variability of the absorbers on short time-scales
are  the fast, relatively-small amplitude fluctuations in the luminosity
of the central source (maximum amplitude of about a factor 2).
As a result of these variability characteristics of the illuminating continuum,
the lack of significant spectral variations may be due to 
the ionization level of the gas reflecting the  {\it time averaged} ionizing luminosity, rather than
very long ionization or recombination times. 
This conclusion therefore depends primarily
on the direct comparison between the low and the high
state spectra, and thus on the S/N in the lower-quality 
spectrum used for the comparison (the high state spectrum). 

To further examine this point, we have calculated the theoretical spectrum 
appropriate for  the high state continuum  under the assumption that the 
absorbing gas responds instantly to changes in the ionizing continuum.
This would be the case if the gas along the line-of-sight gas were of very high density. 
The initial conditions are those assumed for the low-state continuum, and the change in
the SED as described earlier:  a two component X-ray continuum, where the 
long-wavelength component  is assumed to extrapolate down to 40~eV.
The increase in flux between the low- and high states 
was assumed to be a factor 1.5 at 4~\AA\ (see Fig. 3).
Fig. 11 shows two theoretical spectra:
the low-state spectrum (in blue) is the same model shown in Fig. 7,
the high state spectrum (in red) is the result of these new calculations.
The two are clearly different.
In particular the \ion{Si}{7}, \ion{Si}{8} and \ion{Si}{9} lines are
are much weaker for the high-state model.
This is due to the large increase in flux at longest wavelengths, 
with the largest consequences for the lowest
ionization component ($U_{OX}$ has increased by a factor close to 2 
and $U_X$ by a factor of 6.6). 
The variations predicted by this theoretical exercise are much 
larger than those observed (Table 1 and Fig. 4). This illustrates 
the gas did not simply react to the differences in the 
illuminating continuum between the low- and high-states.
An even stronger conclusion is obtained from the fact that 
at long wavelengths the differences observed between the two states 
are much larger than can be  explained by simple variations 
in opacity. 
\begin{figure}
\centerline{\includegraphics[width=10cm]{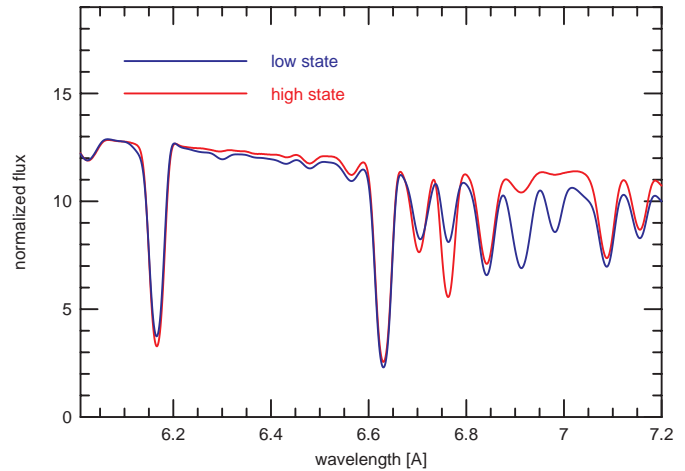}}
\figcaption {
Low state and high state theoretical spectra assuming the observed variable NGC\,3783
continuum and line of sight gas that responds instantly to continuum variations.
The low state model (in blue) is the one described earlier and shown in Fig. 7.
The high state spectrum (in red) assumed the high state SED discussed in the text. Such
a continuum results in much weaker \ion{Si}{8} and \ion{Si}{9} lines.
\label{fig11_label} }
\end{figure}

Given the above findings, we can place limits on the electron density 
($n_e$) and the distance from the central source ($D$) for 
each absorption component. We used the temperatures derived from the models,
average recombination times for the dominant oxygen and silicon ions,  
and assuming there is no response to continuum variations on a time scale of 10 days. 
We find $n_e< 5\times 10^4$ cm$^{-3}$ and $D > 3.2$ pc for the log$(U_{OX})=-2.4$ component,
$n_e< 10^5$ cm$^{-3}$ and $D > 0.63$ pc for the log$(U_{OX})=-1.2$ component ,
 and $n_e< 2.5\times 10^5$ cm$^{-3}$ and $D > 0.18$ pc for the log$(U_{OX})=-0.6$ component.
The corresponding masses are $8.1 \times 10^{3} C_f$, 390$C_f$
 and 64$C_f$  solar masses,
respectively, where $C_f$ is the absorption ($4 \pi$) covering factor. The
mass outflow rate is dominated by the lowest ionization component and is
approximately 
$75 C_f \epsilon v_{\rm outflow}/(500~{\rm km~s^{-1}})$
 solar masses  per year, where $\epsilon$ is the radial filling factor of the flow.
 The very large number suggests
a short duration outburst rather than continuous ejection. We also note that the covering
factor can be small, of order 0.1. 
The above distances are
in agreement with the findings of Behar et al. (2003), who 
have suggested the outflow is located at a distance of a few pc and  is extended beyond 10~pc
based on a comparison between the soft X-ray emission and absorption 
characteristics of the \xmm\ data.

A significant  new aspect of our model is that the product $n_e \times T$
is similar (within a factor $\lesssim$2)  for all three absorption components. 
 This raises the interesting possibility that all the components
are in pressure equilibrium (assuming gas pressure dominates the total pressure, i.e. radiation
pressure and turbulent pressure are not important), 
and they all occupy the same volume of space.
To test this idea, we have calculated thermal stability curves (log$(T)$ vs. log$(U_{OX}/T$)
 for the low-state continuum and our assumed composition (Table 2) under various assumptions.
We have kept the  UV SED unchanged
and varied $\Gamma$(0.1--50 keV) over a wide range. 
We have also investigated the possibility of
a line-of-sight attenuation by the absorbing gas. 
This can be relevant for the situation under study since,
even in the simplest geometry, two of the absorbers do not have a clear view of the 
central radiation source.
Screening by the lowest-ionization component (log$(U_{OX})=-2.4)$ 
is the most important since this gas modifies
the transmitted spectrum much more than the other components (see Fig. 5). 
Three such stability curves are
shown in Fig. 12, along with the location of the three ionization components.
The exact location depends on the details of the assumed SED, the gas metallicity and opacity.
\begin{figure}
\centerline{\includegraphics[width=10cm]{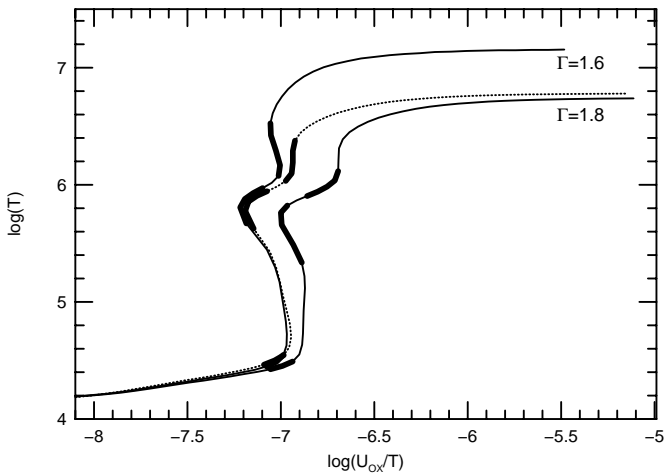}}
\figcaption {
Thermal stability curves for a low density gas exposed to the low-state continuum
of NGC\,3783. The two curves marked by $\Gamma$ are for a bare
continuum with a  0.1--50 keV slope as marked. The dotted line is the stability curve for
a gas exposed to the $\Gamma=1.8$ continuum seen through the log$(U_{OX})=-2.4$
absorber. Thick sections mark the locations of the three absorbers
considered in this
work with the corresponding uncertainties on the ionization parameters.
All ionization components considered in this work
 are situated on thermally stable parts and they all have roughly the same gas pressure.
\label{fig12_label} }
\end{figure}

The interesting feature of the stability diagram is that all radiation fields 
considered here result 
in extended, almost vertical
parts between $T \simeq 3 \times 10^4$\,K and $T \simeq 2 \times 10^6$\,K. 
The curve for $\Gamma=1.6$
has a more extended vertical part since the higher mean energy of the radiation field
results in a Compton temperature.
Screening by the lowest-ionization component in 
NGC\,3783 is indeed important, but does not change the main result that, 
given the uncertainties on the SED and composition, 
all three ionization  components lie on the stable part of the curve. 
We also note that, while here we only considered gas with a single
composition, the details of the stability curve are abundance dependent.
Needless to say, very different curves will be obtained for significantly 
different SEDs.

We note that K03 reach similar conclusions regarding the pressure equilibrium
of their two ionization components.
However, in our model all components are situated on stable branches of the thermal
stability curve (Fig. 12) while according to K03, their high-ionization component is 
situated on the unstable branch (their Fig. 16).  We suspect that
the difference is in part due to the different SED and metalicity assumed by K03,
and in part due to the method they used to calculate the curve.

The existence of an additional stable-region  along the curve of thermal-equilibrium 
(intermediate between the cold and the hot branches) 
where ``warm'' ($few \times 10^5$\,K) can survive is not a new idea.
It has been discussed by Marshall et al. (1993) in their
study of the multi-phase medium in NGC\,1068, and later by many others. 
Reynolds \& Fabian (1995) discussed the stability criteria and
noted the narrow range of ionization parameter allowed
for warm absorbers. Hess et al (1997) discussed the various parameters governing the
stability curve (metallicity and more). 
Komossa and collaborators (e.g. Komossa \& Fink, 1997; Komossa \& Meerschweichen 2001)
calculated many such curves, for different SEDs and metallicities, and discussed the location of
warm absorbers as well as the possible link between the NLR and the X-ray gas.
Krolik \& Kriss (2001) considered 
 multi-phase, warm-absorber winds and noted the various locations of stable, over-heated and
over-cooled gas. Kinkhabwala et al. (2002) re-visited 
the thermal stability issue in NGC\,1068 and showed that a wide range of $U_{OX}$
is needed to account for the observed X-ray emission lines.
Chelouche \& Netzer (2003) investigated the changes in the stability criteria for clouds with
large internal line radiation pressure.

Most of these earlier works focused either on the allowed parts of the curve 
from general stability conditions, or on the location of a certain 
absorber observed in a certain source.  Our work shows that three very different 
aborbers in a source can all occupy extended, stable parts of the curve where
the gas pressure is roughly the same. This means 
there may well be more than three absorbers spread over the vertical branch of
the curves in Fig. 12.   
The three ionization parameters considered here may therefore 
represent some volume-averaged 
properties of the entire cloud ensemble. The column densities are therefore the total
column  densities of a large number of such clouds. 
Moreover, the most-ionized component
may provide the confining medium for the two other ionization components.
Needless to say, real confined outflowing clouds are characterised by a more complicated 
density and pressure structure than with the simplified, constant density clouds considered 
here (e.g. see Chelouche \& Netzer 2001).

The pressure equilibrium between all three 
absorbers in NGC\,3783 suggest they occupy the same general location.
This allows additional constraints to be placed on their density and location. 
Specifically,  we require that the
distance of this region is at least 3.2 pc (the minimum distance of the
lowest ionization component).
However, this zone cannot be
much further away because of total size and volume considerations.
Under the assumed conditions, the highest-ionization parameter component has the
lowest density, and hence the largest dimension. However, its radial extent cannot exceed
about 500 pc (the approximate radius of the narrow-line region), and thus its mean density
cannot fall below about 10 cm$^{-3}$. Pressure equilibrium then dictates  the density of
the lowest-ionization component, and a maximum distance of about 25 pc.
This limitation is only imposed on discrete clouds --- i.e. on the mean
properties of the gas. Real absorbers may not be ``clouds'', they can cover a large
range of distances, and can have properties which depend on location.
The detailed investigation of such models is beyond the scope of the present paper.

\subsection{The emission lines}
There is no indication for flux variation in any of the observed emission
lines. However the measurement uncertainty on these line is rather large 
since the strongest lines are observed at the long wavelengths where the S/N is poor. 
We obtain satisfactory fits for most 
lines by assuming that the emission--line gas has  the same column
density and ionization parameter as the absorbing gas.
The covering factors required for the emitting gas 
are 0.2--0.3 for the log$(U_{OX})=-2.4$ component, 
      0.1--0.2 for the log$(U_{OX})=-1.2$ component, 
and $\sim 0.1$ for the log$(U_{OX})=-0.6$ component.
The large range is due to the uncertainties in the continuum placement and in estimating 
the fraction of the \ion{O}{7} and  \ion{O}{8} emission lines that are being absorbed
by the line-of-sight gas (the model shown in Figs. 7 and 8 assumes that {\it all} the
emitted photons are seen through the absorbers).

A clear shortcoming of the model is the fitting of the intercombination \ion{O}{7}
 line. The intensity predicted for this line is too small by a large factor (2-4).
 A large part of the discrepancy is due to the \ion{O}{6}~$\lambda 21.87$~\AA\ absorption 
line situated close to this emission (Fig. 7). 
This may be related to the \ion{O}{6} problem discussed earlier.
Alternatively it may be a consequence of our assumption
that all emission-line photons are seen through the absorbers.
Indeed, a model that assumes no absorption of the emitted photons by the line-of-sight gas
(not shown here) would give a much better agreement to the \ion{O}{7} absorption complex. 
This would also required a smaller emission covering factor.
Obviously we cannot determine the exact geometry using current observations, 
and the real case may well be intermediate between those two extremes.

We also note on the difference between the absorption covering factor (0.8--1.0) and the
emission covering factor (0.1--0.3). Given the uncertain geometry and the fact that the first is
a line-of-sight covering factor and the secong a globel ($4 \pi$) property, the two are perhaps 
consistent with each other. We did not investigate all possibilities and cannot
comment in detail on special cases like 
 biconical flows or  a torus geometry (e.g. Krolik \& Kriss 2001).

Finally, comparing our method with that of K03 we note that those authors 
do not calculate the emission line flux and profile in a self-consistent way.
Instaed, they add Gaussian shaped emission profiles to their calculated absorption profiles
(e.g. their Fig. 6). This can result in significant differences regarding the line flux
and EW. For example, our default approach assumes that all  emission
lines are seen through the absorber. In such a case, the entire ``blue'' wing of many emission
lines is absorbed. Clearly this affects the EW of the emission lines, but
has almost no effect on the EW of the absorption features.
In contrast, the approach adopted in K03 always results in a decrease
of the EW of the absorption lines.

\subsection{UV absorbers in NGC\,3783}

NGC\,3783 contains time-variable absorbers in the UV that have been described in
various earlier publications (Gabel et al. 2003a and references therein).
Our analysis clearly indicates that at least one X-ray component contains
gas which is of sufficiently low ionization so as to 
produce strong absorption lines of \ion{C}{4}, \ion{N}{5} and \ion{O}{6} in the UV.

The predicted EWs of these UV lines depend on two factors that
have been given little attention in this work: 
the exact shape of the UV continuum, and
covering factor appropriate for the UV absorbers. 
A UV continuum which is softer than that  assumed here can result in
higher levels of ionization for carbon, nitrogen and oxygen, with little effect on
most X-ray lines. This is very important for the \ion{C}{4}$\lambda 1549$~\AA\
and \ion{N}{5}$\lambda 1240$~\AA\ lines, but less so for the UV lines of 
\ion{O}{6} that are directly linked to the \ion{O}{6} column density discussed here.

As for the covering factors, the size of the UV 
continuum source is likely to be  much larger than the size of the central X-ray source. 
Thus the covering factor appropriate in the UV can be different to that 
appropriate in the X-ray.
This will result in saturated UV absorption lines whose EWs are smaller than the
ones indicated by the column densities derived here.
However, our upper limit on the density of the low-ionization component, 
combined with its column density,
suggest a very large line-of-sight dimension ($>10^{17}$ cm).
Assuming lateral dimension of the same size or large, we conclude
that the physical dimension of this component is much larger than the expected
dimension of any likely UV source (e.g. the surface of a thin accretion disk).
A combined analysis of the UV and X-ray results,
taking these such points into consideration,  is in  progress 
            
UV absorbers are also UV emitters, and the low-ionization component can also contribute to
the observed UV emission lines.
Our photoionization calculations show that this contribution is very small.
For example, an emission covering factor of 0.1, similar to the one deduced
from the X-ray emission lines, will produce \ion{O}{6}$\lambda 1035$ line with 
emission EW of about 1\AA.

\subsection{The long wavelength component}
Perhaps the most interesting result of this study is the appearance and
disappearance of the soft excess component. 
This broad-band continuum source was
seen in two of the six observations. On average, these are the 
two observations with the highest luminosity.
An increase in softness ratio with increasing source luminosity is well
known and well documented in a number of AGN
 (e.g. Magdziarz et al., 1998; Chiang et al., 2000; Markowitz \& Edelson 2001).
However we are not aware of a softness ratio increase that is
{\it not correlated} with a short wavelength flux increase.
NGC\,3783
seems to be the first AGN to show this phenomenon, but we suspect that
careful spectroscopic monitoring will reveal the same or
a similar behavior in other sources. This has important consequences to
the continuum production mechanism, as well as for the modeling of the warm
X-ray gas around the center.

We do not know the origin of the soft continuum source. In particular, we
do not have information on its flux at wavelengths 
$\lambda > 30$\AA\  (the source was not in a high-state in
any of the published \xmm\ observations). It may be
related to a broad-band phenomenon (e.g. due  a flaring accretion disk),
or a component covering a narrower band (e.g. a single temperature black body). 
Simple global energy consideration show that this
phenomenon is unrelated to the appearance of broad emission features like those
claimed to be  seen in at least two sources 
and interpreted as due to relativistic disk lines
(Branduardi et al 2001; Mason et al. 2002; but see also Lee et al. 2002).

Finally, we must comment on the possibility of a more complex behavior.
Our analysis is based on the fitting of a single
powerlaw continuum at low-state and the addition of a soft X-ray component during
the high-state. However, an equivalent analysis could be carried out in the reverse
order -- i.e. starting from a pure powerlaw for the  high-state and {\it subtracting}
a continuum component to explain the low-state spectrum. 
However, as argued earlier, the observed spectral 
changes at long wavelengths are too large to be explained by pure opacity variations 
and it is not at all clear what other mechanism could
explain an AGN continuum with flux deficit at long wavelengths.

\section{Conclusions}
Our detailed measurements and analysis of the 900 ks data set of NGC\,3783 lead to
the following results:
\begin{enumerate}
\item
The source fluctuates in luminosity, by a factor $\sim 1.5$, during individual
170 ks observations. The fluctuations are not associated with 
significant spectral variations.
\item
On time scales of  20--120 days, the source exhibits two very
different spectral shapes denoted here as the high- and low-states.
The two are associated with different softness ratios that seem unrelated to the
total X-ray luminosity.
The observed changes in the underlying continuum can be described as due to the
appearance (in the high-state) and disappearance (in the low-state) of a
soft excess component. The origin of this continuum component is not clear.
To the best of our knowledge,  NGC\,3783 is the first AGN to show such a behavior.
\item
The appearance of the soft continuum component can explain all 
spectral
variations observed within the measurement uncertainties. There is no
need to invoke opacity changes between the high- and low-states.
This conclusion depends mostly on the S/N in the high state spectrum.
\item
A combination of three ionization components, each split into two
kinematic components, provides a good description of the
intensity of almost all the absorption lines and bound-free 
edges observed.
The components span a large range of ionization, and have a
total column of about 4$\times 10^{22}$ cm$^{-2}$.
The only real discrepancy between the observed and theoretical spectra 
are for the iron M-shell UTA feature at  16--16.5\AA. This is most likely due to
inadequate dielectronic recombination rates currently available to use
in our calculations.
The largest other uncertainty is in the column density of \ion{O}{6}.
\item
The three generic absorbers discussed in this work have very similar values of
$n_e \times T$. We speculate that the absorbers  may be in pressure equilibrium
with each other, occupying the same volume in the nucleus. This is the first confirmation
of the location of several X-ray absorbers on the vertical
part of the log$(T)$ vs. log$(U_{OX}/T$) stability curve of AGN.
\item
We have obtained thee  lower limits on the gas distance from the center, corresponding
to our three generic absorbers. The limits are 3.2 pc, 0.6 pc and about 0.2 for
the low ionization, intermediate ionization and high ionization absorbers, respectively.
The pressure equilibrium assumption implies distances in the range 3--25~pc.
\end{enumerate}

\acknowledgements
This work is supported by the Israel Science Foundation grant 545/00,
H.N. thanks S. Kahn and the astrophysics group at Columbia university for hospitality
and support during a summer visit in 2002.
E.B. was supported by the Yigal-Alon Fellowship and by the GIF Foundation under 
grant \#2028-1093.7/2001.
We gratefully acknowledge the financial support of CXC grant GO1-2103
(S. K., W. N. B., I. M. G.), NASA LTSA grant NAG~5-13035 (W. N. B.) 
and the Alfred P. Sloan Foundation (W. N. B.),

\end{document}